\documentclass[a4paper,11pt]{article}
\pdfoutput=1
\usepackage{array}
\usepackage{float}
\usepackage{amsmath}
\usepackage{amsfonts}
\usepackage{amssymb}
\usepackage{latexsym}
\usepackage{graphicx}
\usepackage{mathrsfs}
\usepackage{fancyhdr}
\usepackage{sectsty}
\usepackage[longnamesfirst]{natbib}
\usepackage{theorem}
\usepackage{url}
\usepackage{rotating}
\usepackage[margin=1in]{geometry}

\newtheorem{theorem}{Theorem}

{
\theorembodyfont{\upshape}
\theoremheaderfont{\scshape}

}
\newtheorem{assumption}{Assumption}

\newtheorem{corollary}{Corollary}

\newtheorem{example}{{\sc Example}}
{\theorembodyfont{\upshape}
\newtheorem{remark}{Remark}[section]
}

\newtheorem{lemma}{{\sc Lemma}}

\newenvironment{proof}[1][Proof]{\bigskip \noindent \textbf{#1:} }{\  \rule{0.5em}{0.5em}}

\theoremheaderfont{\bfseries}
\theorembodyfont{\upshape}

\sectionfont{\centering \sc}
\subsectionfont{\centering \sc}
\subsubsectionfont{\centering \sc}
\allsectionsfont{\centering \sc}
\allowdisplaybreaks
\renewcommand{\cite}{\citet}

\begin{document}

\begin{center}
{\Large BOOTSTRAPPING NON-STATIONARY STOCHASTIC VOLATILITY}

\mbox{}

\vspace{-0.1cm}

\renewcommand{\thefootnote}{}
\footnote{
\hspace{-7.2mm}
$^{a}$Amsterdam School of Economics and Tinbergen Institute, University of Amsterdam, The Netherlands\\
$^{b}$Department of Economics, University of Bologna, Italy\\
$^{c}$Department of Economics, Exeter Business School, UK\\
$^{d}$Institut d'An\`{a}lisi Econ\`{o}mica (CSIC), Spain\\
$^{e}$Department of Economics, University of Copenhagen, Denmark\\
Correspondence to: Giuseppe Cavaliere, Department of Economics, University of Bologna,
Piazza Scaravilli 2, I-40126 Bologna, Italy. email: giuseppe.cavaliere@unibo.it.}
\addtocounter{footnote}{-1}
\renewcommand{\thefootnote}{\arabic{footnote}}%

{\normalsize \vspace{0.1cm} }

\textsc{H.\ Peter Boswijk}$^{a}$\textsc{, Giuseppe Cavaliere}$^{b,c}$\textsc{%
, Iliyan Georgiev}$^{b,d}$\textsc{\ }\\[0pt]
\textsc{and Anders Rahbek}{\large $^{e}$}
\end{center}

\par\begingroup\leftskip=1 cm
\rightskip=1 cm
\small%

\begin{center}
\textsc{Abstract\vspace{-0.15cm}}
\end{center}

In this paper we investigate to what extent the bootstrap can be applied to
conditional mean models, such as regression or time series models, when the
volatility of the innovations is random and possibly non-stationary. In
fact, the volatility of many economic and financial time series displays
persistent changes and possible non-stationarity. However, the theory of the
bootstrap for such models has focused on deterministic changes of the
unconditional variance and little is known about the performance and the
validity of the bootstrap when the volatility is driven by a non-stationary
stochastic process. This includes near-integrated exogenous volatility
processes as well as near-integrated GARCH processes, where the conditional
variance has a diffusion limit; a further important example is the case
where volatility exhibits infrequent jumps. This paper fills this gap in the
literature by developing conditions for bootstrap validity in time series
and regression models with non-stationary, stochastic volatility. We show
that in such cases the distribution of bootstrap statistics (conditional on
the data) is random in the limit. Consequently, the conventional approaches
to proofs of bootstrap consistency, based on the notion of weak convergence
in probability of the bootstrap statistic, fail to deliver the required
validity results. Instead, we use the concept of `weak convergence in
distribution' to develop and establish novel conditions for validity of the
wild bootstrap, conditional on the volatility process. We apply our results
to several testing problems in the presence of non-stationary stochastic
volatility, including testing in a location model, testing for structural
change using CUSUM-type functionals, and testing for a unit root in
autoregressive models. Importantly, we work under sufficient conditions for
bootstrap validity that include the absence of statistical leverage effects,
i.e., correlation between the error process and its future conditional
variance. The results of the paper are illustrated using Monte Carlo
simulations, which indicate that a wild bootstrap approach leads to size
control even in small samples.

\bigskip

\noindent\textsc{Keywords: }Bootstrap; Non-stationary stochastic volatility;
Random limit measures; Weak convergence in Distribution.

\bigskip

\noindent\textsc{JEL Classification}: C32.

\bigskip

\par\endgroup\normalsize%

\section{Introduction}

\textsc{In this paper} we consider bootstrap and asymptotic inference on the
conditional mean in econometric time series models when the (conditional)
volatility is allowed to show a large degree of persistence due to possible
permanent and stochastic changes, reflecting the well established fact that
volatility in many economic and financial time series displays high
persistence, and covariance non-$\frac{{}}{{}}$stationarity.

Earlier references in macroeconomics include \cite{KN99} and \cite{MP00},
who find evidence of an (unanticipated) structural change in the volatility
of US GDP growth rates. Evidence of changes in the unconditional volatility
appear in many key time series, such as aggregate consumption and income, in
interest rate data and in nominal and real price variables; see \cite{SvD04}%
. Evidence on changes in the long-run component of volatility in stock and
currency markets are initially reported in \cite{LP94} and \cite{H95}, who
show that when stochastic volatility [SV] models are taken to the data, the
largest autoregressive root in the SV process is so close to one that the
assumption of stationary volatility seems to be at odds with the data.
Similarly, it is a well-known stylized fact that GARCH models fit to stock
market returns display parameter estimates which reflect high persistence as
they (nearly) violate covariance stationarity conditions (often referred to
as \textquotedblleft near-integrated GARCH\textquotedblright ), and that
such parameters are smaller when a slowly-varying long run component is
accounted for in the model, see \cite{ER08}. \cite{HLST16} list a number of
empirical studies that have found strong evidence of structural breaks in
the unconditional variance of asset returns, with break dates driven by
major financial and macroeconomic crises. Such (possibly random) volatility
shifts are known to affect the asymptotic properties of estimators of the
parameters of models for the conditional mean; see \cite{CT07}, \cite{XP08}
and, for multivariate models, Cavaliere, Rahbek and Taylor (2010a;b) and
Boswijk, Cavaliere, Rahbek and Taylor (2016).

In the framework of a conditional mean, or a general (stationary, or
non-stationary) regression type model, the \emph{wild bootstrap} is an
important tool to deliver consistent estimation of the asymptotic
distributions of test statistics or parameter estimators. The wild bootstrap
allows in particular to track changes in the quadratic variation of an
econometric model by simply mimicking the (unknown) volatility dynamics
through the squared model residuals, see \citet{GK04, GK07} for applications
to stationary time series models and \citet{CRT10a,CRT10b} for
non-stationary multivariate models.

Consider the simple case where the volatility, say $\sigma _{t}$, can be
approximated by a \emph{non-stochastic} element of the space $%
\mathscr{D}[0,1]$ of c\`{a}dl\`{a}g functions on $[0,1]$, such that $\sigma
_{t}=\sigma (t/n)$ ($t=1,\ldots ,n$, $n$ denoting the sample size) with $%
\sigma \in \mathscr{D}[0,1]$. Simple special cases are a single volatility
break at time $\lfloor n\tau \rfloor $ (with $\lfloor \cdot \rfloor $
denoting the (floor) integer value), for some $\tau \in (0,1)$, as given by
(with $\mathbb{I}_{A}(\cdot )$ denoting the indicator function of the set $A$%
) 
\begin{equation*}
\sigma (u):=\sigma _{A}+(\sigma _{B}-\sigma _{A})\mathbb{I}_{[\tau ,1]}(u),%
\text{ for }u\in \left[ 0,1\right] \text{,}
\end{equation*}%
where $\sigma_A >0$ and $\sigma_B >0$; or the case of trending volatility,%
\begin{equation*}
\sigma (u):=\sigma _{A}+(\sigma _{B}-\sigma _{A})u^{\delta },
\end{equation*}%
where $\sigma_A >0$, $\sigma_B >0$ and $\delta >0$.
A classic wild bootstrap, based on the resampling scheme $%
\varepsilon _{t}^{\ast }=\hat{\varepsilon}_{t}w_{t}^{\ast }$, where the $%
\hat{\varepsilon}_{t}$'s are the estimated residuals from the regression
model and the $w_{t}^{\ast }$'s are i.i.d.\ $(0,1)$ bootstrap shocks,
independent of the original sample, is in general able to track the
volatility path (in terms of quadratic variation) of the original data,
without any assumption on the initial values of the volatility process as,
loosely speaking, under standard assumptions%
\begin{equation*}
n^{-1}\sum_{t=1}^{\lfloor nu\rfloor }(\varepsilon _{t}^{\ast
})^{2}=n^{-1}\sum_{t=1}^{\lfloor nu\rfloor }\hat{\varepsilon}%
_{t}^{2}(w_{t}^{\ast })^{2}\approx n^{-1}\sum_{t=1}^{\lfloor nu\rfloor
}\varepsilon _{t}^{2}+o_{p}(1)=\int_{0}^{u}\sigma (s)^{2}\mathsf{d}s+o_{p}(1)%
\text{.}
\end{equation*}%
Existing theory of the bootstrap mainly focuses on such deterministic
changes of the unconditional variance and little is known about the
performance and the validity\footnote{%
Throughout the paper, with `validity' of the bootstrap we mean that the
associated bootstrap tests control size asymptotically.
With `consistency of the bootstrap test' we mean that the (bootstrap) test rejects with
probability tending to one under the alternative.} of the bootstrap when the
volatility is driven by a high-persistent, or (second order) non-stationary
stochastic process. This includes leading key cases such as near-integrated
exogenous volatility processes (as analyzed by \citealp{H95}), as well as
near-integrated GARCH processes, where the conditional variance has a
diffusion limit \citep{N90}.

This paper fills this gap in the literature by developing conditions for
bootstrap validity and consistency of the associated bootstrap tests in
regression and time series models with persistent stochastic volatility.
That is, we replace the deterministic volatility assumption by allowing that
volatility is the realization of a (non-stationary) stochastic process $%
\sigma _{t}$; specifically, we derive results under the general assumption
that, for the c\`{a}dl\`{a}g version of the volatility, it holds that%
\begin{equation}
\sigma _{\lfloor un+1\rfloor }\overset{w}{\rightarrow }\sigma (u)\text{ for }%
u\in \left[ 0,1\right] \text{,}  \label{eq conv of sigma_t to stoch limit}
\end{equation}%
where $\sigma $ is some random element in $\mathscr{D}[0,1]$.

The analysis of the bootstrap under a weak convergence assumption like (\ref%
{eq conv of sigma_t to stoch limit}) is not straightforward. As we show in
the paper, a key fact under non-stationary stochastic volatility is that the
distribution of bootstrap statistics (conditional on the data), rather than
converging to the unconditional distribution of the statistic of interest,
converges weakly to a \emph{random }limit. By this we mean that the
distribution function of the bootstrap statistic (conditional on the data)
is stochastic not only for finite sample sizes $n$, but also in the limit as 
$n\rightarrow \infty $. Consequently, the conventional approach, based on
the notion of weak convergence in probability of the bootstrap statistic to
the limiting distribution of the original statistic (which is obviously
non-stochastic), fails to deliver the required result of validity of the
bootstrap. This problem is not new in the bootstrap literature, as it
appears in various areas of application of the bootstrap; for example, in
models with infinite variance innovations \citep{K89} and in autoregressive
models with unit roots \citep{B91,CNR15}.

Specifically, in this paper we analyze the wild bootstrap under
(non-stationary) stochastic volatility by adopting a new approach to assess
bootstrap validity under random limit bootstrap measures. Thus, rather than
focusing on the usual weak convergence in probability of the bootstrap
conditional distribution, we make use of the concept of weak convergence in
distribution (see \citealp{CG20}, for a general introduction) to develop
novel conditions for validity of the wild bootstrap, conditional on the
volatility process. This allows us to establish that, although the presence
of a \emph{random} limiting distribution for the bootstrap statistic makes
the bootstrap unable to estimate the unconditional distribution of the
statistic of interest, the bootstrap can still deliver hypothesis tests with
the desired size. In particular, we do this by establishing that the
high-level conditions for bootstrap validity in \cite{CG20} can be shown to
hold for a large class of models with stochastic volatility, including the
aforementioned near-integrated GARCH model and the non-stationary stochastic
volatility model. We do so by showing new weak convergence results
conditional on volatility paths.

To illustrate our new approach and its applicability, we apply our results
to three leading testing problems in the presence of non-stationary
stochastic volatility, including testing a hypothesis on the location of a
time series, testing for a unit root and testing for stability of the
conditional mean using CUSUM-type statistics. These illustrative examples
can easily be extended to cover more general cases, such as cointegration
(as in \citealp{CRT10a}, \citealp{CNR15} and \citealp{BCRT16}) with
multivariate stochastic volatility, or multivariate stability tests (see %
\citealp{P06} and \citealp{CP19}). Importantly, for all examples we show
that conditions for conditional wild bootstrap validity include the absence
of statistical leverage effects, i.e.\ correlation between the error process
and its future conditional variance. The results of the paper are
illustrated using Monte Carlo simulations, which indicate that under the
conditions developed in our paper, the wild bootstrap leads to excellent
size control even in small samples.

\subsection*{Structure of the paper}

The structure of the paper is the following. In Section \ref{Sec Model} we
introduce the reference data generating process and our main assumptions, in
particular on the volatility. Here we also introduce three examples which
are used throughout the paper to illustrate the main results. We also derive
the reference limit distribution for non-bootstrap statistics under
non-stationary volatility. In Section \ref{Sec invalidity of the BS} we
introduce the main (wild) bootstrap algorithm. We show that when volatility
is non-stationary, the bootstrap fails to mimic the asymptotic distribution
of the corresponding statistics, and hence it is not valid in the usual
sense. In Section \ref{sec BS conditional validity} we discuss the wild
bootstrap and prove, under proper assumptions, validity conditionally on the
volatility path, as well as consistency of the bootstrap tests under the
alternative. We first introduce in Section \ref{sec wcd and bootstrap
validity} the concept of weak convergence in distribution and discuss how to
prove validity of the bootstrap in the presence of random limit bootstrap
distributions, as it happens here under non-stationary volatility. Then in
Section \ref{sec main result} we provide our main results under the required
additional conditions on the original data. In Section \ref{sec
examples-validity} we apply our results about validity of the bootstrap in
our three applications. Finally, in Section \ref{sec behaviour under
alternative} we discuss consistency of the bootstrap tests under the
alternative hypothesis. Results from a Monte Carlo study on the finite
sample behavior of the bootstrap tests are reported in Section \ref{Sec MC}.
Section \ref{Sec conclusions} concludes. All proofs are reported in the
online appendix.

\subsection*{Notation}

The following (standard) notation is used throughout. With $x:=y$ ($y=:x$)
we mean that $x$ is defined by $y$ ($y$ defined by $x$). For any $q\in 
\mathbb{R}$ ($\mathbb{R}$ denoting the set of real numbers), $\lfloor
q\rfloor $ denotes the integer part of $q$. For random elements $(X_{n},X)$
of a metric space, weak convergence of $X_{n}$ to $X$ is denoted by $X_{n}%
\overset{w}{\rightarrow }X$. Also, $\overset{d}{=}$ denotes equality in
distribution. We use $P^{\ast }$, $E^{\ast }$ and $V^{\ast }$ respectively
to denote probability, expectation and variance, conditional on the original
sample. With $\overset{w^{\ast }}{\rightarrow }_{p}$ we denote weak
convergence in probability; in particular, for random varables $(X_{n}^{\ast
},X)$, the notation $X_{n}^{\ast }\overset{w^{\ast }}{\rightarrow }_{p}X$
means that, as the sample size $n$ diverges, the cumulative distribution
function [cdf] $G_{n}^{\ast }$ of $X_{n}$, conditional on the original data,
converges in probability to the cdf $G$ of $X$, at all continuity points of $%
G$. For a given sequence $X_{n}^{\ast }$ of random elements, computed from
the bootstrap data, $X_{n}^{\ast }-X=o_{p}^{\ast }(1)$, in probability, or $%
X_{n}^{\ast }\overset{p^{\ast }}{\rightarrow }_{p}X,$ means that for any $%
\epsilon >0$, $P^{\ast }(||X_{n}^{\ast }-X||>\epsilon )\overset{p}{%
\rightarrow }0$, as $n\rightarrow \infty $. Similarly, $X_{n}^{\ast
}=O_{p}^{\ast }(1)$, in probability, means that, for every $\epsilon >0$,
there exists a constant $M>0$ such that, for all large $n$, $P(P^{\ast
}(||X_{n}^{\ast }||>M)<\epsilon )$ is arbitrarily close to one.
Weak convergence in distribution and the related notation are introduced in Section \ref{sec wcd and bootstrap validity}.
The Skorokhod spaces of c\`{a}dl\`{a}g functions $[0,1]\rightarrow \mathbb{R}%
^{m\times n}$ and $[0,1]\rightarrow \mathbb{R}^{n}$ are denoted by $%
\mathscr{D}{}_{m\times n}[0,1]$ and $\mathscr{D}{}_{n}[0,1]$, respectively;
for the latter, when $n=1$ the subscript is suppressed. The Skorokhod space
of c\`{a}dl\`{a}g functions $\mathbb{R}\rightarrow \mathbb{R}$ is denoted by 
$\mathscr{D}{}(\mathbb{R})$.

\section{Set-up and preliminaries}

\label{Sec Model}

In this section we introduce our reference class of models for the
conditional mean under stochastic volatility as well as the (test)
statistics of interest. In Section \ref{sec testing problems} we focus on
statistics which can be expressed (at least when the associated null
hypothesis holds true) as functionals of the partial sum of the innovations
and of the partial sum of the squared innovations. To illustrate ideas, we
consider three simple univariate cases (which can easily be extended to
multivariate cases) throughout: (i) testing a hypothesis on the mean in a
simple location model; (ii) CUSUM\ testing for parameter constancy in a
location model; (iii)\ testing for an autoregressive unit root in an AR(1)
model. The main assumption on the volatility --- which, \emph{inter alia},
allows for non-stationary stochastic volatility or near-integrated GARCH\
dynamics --- is discussed next in Section \ref{sec intro to NSV}. Under the
assumptions in Sections \ref{sec testing problems} and \ref{sec intro to NSV}%
, the asymptotic (null) distributions can be derived. We do this in Section %
\ref{sec standard asy}, where we show that the limiting distribution can be
expressed in terms of a continuous martingale and its quadratic variation
process. The implications of these results on bootstrap inference and
hypothesis testing are the focus of the main Sections \ref{Sec invalidity of
the BS} and \ref{sec BS conditional validity}.

\subsection{Model and hypotheses of interest}

\label{sec testing problems}

We are concerned with inference and hypothesis testing on the regression
parameters of a heteroskedastic time series regression model in a triangular
array form: 
\begin{equation}
y_{n,t}=\beta ^{\prime }x_{n,t}+\varepsilon _{n,t},\qquad t=1,\ldots ,n;%
\text{ }n=1,2,\ldots  \label{regr}
\end{equation}%
where $\varepsilon _{n,t}$ is a martingale difference sequence (mds)
relative to a suitable filtration $\mathcal{F}_{n,t}$, with conditional
variance $\sigma _{n,t}^{2}=E(\varepsilon _{n,t}^{2}|\mathcal{F}_{n,t-1})$.
To simplify notation, unless strictly required we simply write (\ref{regr})
as $y_{t}=\beta ^{\prime }x_{t}+\varepsilon _{t}$, with $\sigma
_{t}^{2}:=E(\varepsilon _{t}^{2}|\mathcal{F}_{t-1})$.

Inference focuses on test statistics, which we assume can be expressed (at
least under the null hypothesis) as functionals of partial sum processes in
terms of the innovations and squared innovations 
\begin{equation}
\left( M_{n}(u),U_{n}(u)\right) :=\left(
n^{-1/2}\textstyle \sum_{t=1}^{\lfloor nu\rfloor }\varepsilon
_{t},n^{-1}\sum_{t=1}^{\lfloor nu\rfloor }\varepsilon
_{t}^{2}\right) ,\qquad u\in \lbrack 0,1].  \label{Mn}
\end{equation}%
as is the case for many testing problems, see also the discussion and
examples below.

Defining $\mathcal{F}_{n}(u):=\mathcal{F}_{\lfloor nu\rfloor }$, the mds
assumption implies that $\left\{ M_{n}(u),\mathcal{F}_{n}(u)\right\} _{u\in
\lbrack 0,1]}$ is a martingale for all $n$, and $U_{n}(u)$ is its quadratic
variation process, i.e., 
\begin{equation}
U_{n}(u)=[M_{n}](u)=\sum_{t=1}^{\lfloor nu\rfloor }\left( M_{n}\left( \frac{t%
}{n}\right) -M_{n}\left( \frac{t-1}{n}\right) \right) ^{2},\qquad u\in
\lbrack 0,1].  \label{Vn}
\end{equation}%
Throughout it will also be useful to define the predictable quadratic
variation or angle bracket process (see \citealp{JS03}): 
\begin{equation}
V_{n}(u):=\langle M_{n}\rangle (u)=n^{-1}\textstyle \sum_{t=1}^{\lfloor
nu\rfloor }\sigma _{t}^{2},\qquad u\in \lbrack 0,1],  \label{Un}
\end{equation}%
with the defining property that $\left\{ M_{n}^{2}(u)-\langle M_{n}\rangle
(u),\mathcal{F}_{n}(u)\right\} _{u\in \lbrack 0,1]}$ is a martingale.

The following three testing problems are discussed in the paper. These are
all special cases of (\ref{regr}) where the statistic of interest is indeed
a functional of $\left( M_{n},U_{n}\right) $.

\bigskip

\noindent \textsc{Example 1 (testing in a location model). }Consider the
location model $y_{t}=\theta +\varepsilon _{t}$, which is trivially obtained
from (\ref{regr}) by setting $\beta =\theta $ and $x_{t}=1$. The true
location parameter is denoted by $\theta _{0}$. Suppose that interest is in
testing the simple null hypothesis $\theta =\bar{\theta}$. Then, one can
consider the test statistic $S_{n}:=\sqrt{n}(\overline{y}_{n}-\bar{\theta})$%
, where $\overline{y}_{n}=n^{-1}\sum_{t=1}^{n}y_{t}$, or, alternatively, its
studentized version $T_{n}:=\sqrt{n}(\overline{y}_{n}-\bar{\theta})/s_{n}$,
with $s_{n}^{2}=n^{-1}\sum_{t=1}^{n}(y_{t}-\bar{y}_{n})^{2}$. It is not
difficult to see that, under the null hypothesis, it holds that $S_{n}$ and $%
T_{n}$ can be expressed in terms of $M_{n}$ and $U_{n}$ defined in (\ref{Mn}%
) as%
\begin{equation*}
S_{n}=\sqrt{n}(\overline{y}_{n}-\bar{\theta})=M_{n}(1),\qquad T_{n}=\sqrt{n}%
\frac{(\overline{y}_{n}-\bar{\theta})}{s_{n}}=\frac{M_{n}(1)}{\sqrt{%
U_{n}(1)-n^{-1}M_{n}(1)^{2}}}.
\end{equation*}%
If $s_{n}$ is constructed with the null imposed, i.e. $s_{n}^{2}=n^{-1}%
\sum_{t=1}^{n}(y_{t}-\bar{\theta})^{2}$, then under the null hypothesis $T_{n}$ simplifies to%
\begin{equation*}
T_{n}=\frac{M_{n}(1)}{\sqrt{U_{n}(1)}}\text{.}
\end{equation*}

\bigskip

\noindent \textsc{Example 2 (CUSUM test in a location model). }Consider the
time-varying location model $y_{t}=\theta _{t}+\varepsilon _{t}$, and
suppose that interest is in testing the null hypothesis of a constant
location parameter, i.e. $\mathsf{H}_{0}:\theta _{t}=\theta _{1}$, $%
t=2,\ldots ,n$. A standard CUSUM test can be constructed by considering the
statistic (see e.g.\ \citealp{DP08}, and the references therein) 
\begin{equation*}
CS_{n}:=\frac{1}{n^{1/2}}\max_{t=1,\ldots ,n}\left\vert \sum_{i=1}^{t}(y_{i}-%
\bar{y}_{n})\right\vert ,
\end{equation*}%
or its studentized version, 
\begin{equation*}
CT_{n}:=\frac{1}{s_{n}n^{1/2}}\max_{t=1,\ldots ,n}\left\vert
\sum_{i=1}^{t}(y_{i}-\bar{y}_{n})\right\vert ,
\end{equation*}%
which, as in Example 1, reduce to 
\begin{equation*}
CS_{n}=\sup_{u\in \lbrack 0,1]}|M_{n}(u)-uM_{n}(1)|\text{, }\hspace{1cm}%
CT_{n}=\frac{\sup_{u\in \lbrack 0,1]}|M_{n}(u)-uM_{n}(1)|}{\sqrt{%
U_{n}(1)-n^{-1}M_{n}^{2}(1)}}
\end{equation*}%
under $\mathsf{H}_{0}$.\bigskip

\noindent \textsc{Example 3 (Testing for a unit root}) Consider the
first-order autoregression $y_{t}=(1+\theta )y_{t-1}+\varepsilon _{t}$, with 
$y_{0}=0$ (which again follows from (\ref{regr}) by setting $\beta =1+\theta 
$ and $x_{t}=y_{t-1}$). A test of the unit root hypothesis $\theta =0$ can
be based on the Dickey-Fuller `coefficient' statistic $R_{n}:=n\hat{\theta }%
_{n}$, where $\hat{\theta }_{n}=\sum_{t=1}^{n}y_{t-1}\Delta
y_{t}/\sum_{t=1}^{n}y_{t-1}^{2}$ is the least-squares estimator from the
regression of $\Delta y_{t}$ on $y_{t-1}$. Under the null hypothesis, $\hat{%
\theta }_{n}=\sum_{t=1}^{n}\varepsilon _{t}(\sum_{i=1}^{t-1}\varepsilon
_{i})/\sum_{t=1}^{n}(\sum_{i=1}^{t-1}\varepsilon _{i})^{2}$ and the test
statistic may be expressed as 
\begin{equation*}
R_{n}=\frac{\int_{0}^{1}M_{n}(u)\mathsf{d}M_{n}(u)}{\int_{0}^{1}M_{n}^{2}(u) 
\mathsf{d}u}=\frac{\frac{1}{2}\left( M_{n}^{2}(1) -U_{n}(1) \right) }{%
\int_{0}^{1}M_{n}^{2}(u) \mathsf{d}u} .
\end{equation*}%
If the test is based on the Dickey-Fuller `ratio' statistic $W_{n}:=\hat{%
\theta }_{n}(s_{n}/(\sum_{t=1}^{n}y_{t-1}^{2}))^{-1/2}$, where $%
s_{n}^{2}:=n^{-1}\sum_{t=1}^{n}(\Delta y_{t}-\hat{\theta}y_{t-1})^{2}$, then 
\begin{equation*}
W_{n}=\frac{\int_{0}^{1}M_{n}(u)\mathsf{d}M_{n}(u)}{\sqrt{%
\int_{0}^{1}M_{n}^{2}(u) \mathsf{d}u}}\frac{1}{ \sqrt{ U_{n}(1)
-n^{-1}(\int_{0}^{1}M_{n}(u)\mathsf{d}M_{n}(u))^{2}/\int_{0}^{1}M_{n}^{2}(u) 
\mathsf{d}u }}
\end{equation*}%
under the null hypothesis.$\hfill \square $

\bigskip

Some remarks are in order.

\begin{remark}
Although for fixed $n$, $U_{n}$ can be determined from $M_{n}$ as seen in (%
\ref{Vn}), it does not define a continuous function $h:\mathscr{D}[0,1]%
\rightarrow \mathscr{D}[0,1]$. Therefore, limit results for $U_{n}$ cannot
be obtained from weak convergence of $M_{n}$ together with the continuous
mapping theorem [CMT]. Joint weak convergence of $\left( M_{n},U_{n}\right) $
is required to obtain the asymptotic null distribution of the statistics $%
T_{n}$, $CT_{n}$, $R_{n}$ and $W_{n}$. This is related to the well-known
fact that weak convergence of $\int_{0}^{1}M_{n}\mathsf{d}M_{n}$ to the
stochastic integral $\int_{0}^{1}M\mathsf{d}M$ does not follow from $M_{n}%
\overset{w}{\rightarrow }M$ and the CMT; see e.g.\ \cite{CW88}.
\end{remark}

\begin{remark}
The above examples involve single-parameter models. In more general testing
situations, such as testing for a unit root in higher-order autoregressive
models, the statistic of interest may be written as a functional of $\left(
M_{n},U_{n}\right) $ plus an asymptotically negligible term. The theory
developed in this paper can be extended to cover such cases.$\hfill \square $
\end{remark}

\subsection{Non-stationary stochastic volatility}

\label{sec intro to NSV}We now introduce our basic hypotheses on the dynamic
behavior of the conditional volatility $\sigma _{t}^{2}$ of the shocks $%
\varepsilon _{t}$. More specifically, we will allow volatility to be a
persistent stochastic process, with a stochastic volatility weak limit, as
formulated in the next two assumptions. These are in the spirit of the
seminal paper by \cite{H95}, who considers conditional variances driven by
nearly-integrated autoregressive shocks, although we do not constrain the
behavior of the conditional variance to be of the autoregressive type.

\begin{assumption}
\label{ass:mds}In (\ref{regr}), we have $\varepsilon _{t}=\sigma _{t}z_{t}$,
where $z_{t}$ is a martingale difference sequence relative to $\mathcal{F}%
_{t}=\sigma (\left\{ z_{i}\right\} _{i=1}^{t},\left\{ \sigma _{i}\right\}
_{i=1}^{t+1})$, satisfying $E(z_{t}^{2}|\mathcal{F}_{t-1})=1$.
\end{assumption}

\noindent Define now the $\mathscr{D}{}[0,1]$ version of the partial sum of
the $z_{t}$'s as $B_{z,n}(u):=n^{-1/2}\sum_{t=1}^{\lfloor nu\rfloor }z_{t}$, 
$u\in \lbrack 0,1]$, and $\sigma _{n}(u)$, $u\in \lbrack 0,1]$, the $%
\mathscr{D}{}[0,1]$ version of $\sigma _{t}$, as:%
\begin{equation}
\sigma _{n}(u):=\sigma _{\lfloor nu+1\rfloor }\text{, for }u\in \lbrack 0,1)%
\text{, }  \label{eq sigma_n(u) and B_z,n(u)}
\end{equation}%
with $\sigma _{n}(1):=\sigma _{n}$. In the following, the process $\sigma
_{n}(u)$, $u\in \lbrack 0,1]$, will be referred to as $\sigma _{n}$; this
should not be confused with $\sigma _{t}$ for $t=n$, as also will be clear
from the context where the notation is used.

\begin{assumption}
\label{ass:SV}As $n\rightarrow \infty $, $\left( \sigma _{n}(u) ,B_{z,n}(u)
\right) \overset{w}{\rightarrow }\left( \sigma (u) ,B_{z}(u) \right) $ in $%
\mathscr{D}_{2}\left[ 0,1\right] $, where $\sigma \in \mathscr{D}{}[0,1]$ is
a stochastic process satisfying $\inf_{u\in \lbrack 0,1]}\sigma (u)>0~%
\mathrm{a.s.}$, and $B_{z}$ is a standard Brownian motion on $[0,1]$.
\end{assumption}

\noindent While the convergence of the partial sum $B_{z,n}\in %
\mathscr{D}{}[0,1]$ is standard, the requirement on $\sigma _{n}\in %
\mathscr{D}{}[0,1]$ is not. More specifically, this assumption requires the
conditional variance process $\{\sigma _{t}^{2}\}_{t=1}^{n}$ to be
persistent enough such that its behavior can be approximated by an element
of the space of c\`{a}dl\`{a}g functions $\mathscr{D}{}[0,1]$. No higher
moment conditions are required on $\{\sigma _{t}^{2}\}_{t=1}^{n}$ or $%
\{\varepsilon _{t}\}_{t=1}^{n}$. Some examples of processes satisfying
Assumptions \ref{ass:mds} and \ref{ass:SV} are presented next. These will be
analyzed in detail throughout the paper; for additional cases and
discussions see e.g.\ \citet{CT09}.

\renewcommand{\theexample}{V.\arabic{example}}

\begin{example}[Stochastic volatility]
\label{ex:SV}Let $\sigma _{t}^{2}$ be generated by 
\begin{equation*}
\log \sigma _{t}^{2}=\phi _{n}\log \sigma _{t-1}^{2}+(1-\phi _{n})\log \bar{%
\sigma}^{2}+n^{-1/2}\eta _{t-1},\qquad t=1,2,\ldots
\end{equation*}%
where $\sigma _{0}^{2}=\bar{\sigma}^{2}$ for some $\bar{\sigma}>0$, $\phi
_{n}=e^{-\kappa /n}$ for some $\kappa \geq 0$, and $\eta _{t}\sim \mathrm{%
i.i.d.}~N(0,\sigma _{\eta }^{2})$, independent of $z_{t}\sim \mathrm{i.i.d.}%
~N(0,1)$. Then $\left( \sigma _{n},B_{z,n}\right) \overset{w}{\rightarrow }%
\left( \sigma ,B_{z}\right) $ in $\mathscr{D}_{2}[0,1]$, where:%
\begin{equation*}
\mathsf{d}\log \sigma ^{2}(u)=\kappa (\log \sigma ^{2}(u)-\log \bar{\sigma}%
^{2})\mathsf{d}u+\sigma _{\eta }\mathsf{d}B_{\eta }(s),\qquad u\in \lbrack
0,1],
\end{equation*}%
with $\sigma ^{2}(0)=\bar{\sigma}^{2}$, and where $(B_{\eta },B_{z})$ is a
bivariate standard Brownian motion.
\end{example}

\begin{example}[Near-integrated GARCH]
\label{ex:GARCH}Consider the case where $\sigma _{t}^{2}$ is generated by
the standard GARCH\ recursion%
\begin{equation*}
\sigma _{t}^{2}=\omega _{n}+\alpha _{n}\varepsilon _{t-1}^{2}+\beta
_{n}\sigma _{t-1}^{2}=\omega _{n}+\alpha _{n}\sigma
_{t-1}^{2}z_{t-1}^{2}+\beta _{n}\sigma _{t-1}^{2},\qquad t=1,2,\ldots
\end{equation*}%
where $\sigma _{0}^{2}=\bar{\sigma}^{2}$ for some $\bar{\sigma}>0$, $\alpha
_{n}+\beta _{n}=1-n^{-1}\kappa $ for some $\kappa \geq 0$, $\alpha
_{n}=(2n)^{-1/2}\sigma _{\eta }$ for some $\sigma _{\eta }>0$, $\omega
_{n}=n^{-1}\bar{\sigma}^{2}\kappa $ and $z_{t}\sim \mathrm{i.i.d.}~N(0,1)$.
Then it follows from \citet{N90} that $\left( \sigma _{n},B_{z,n}\right) 
\overset{w}{\rightarrow }\left( \sigma ,B_{z}\right) $ in $\mathscr{D}%
_{2}[0,1]$, where 
\begin{equation*}
\mathsf{d}\sigma ^{2}(u)=\kappa (\sigma ^{2}(u)-\bar{\sigma}^{2})\mathsf{d}%
u+\sigma _{\eta }\sigma ^{2}(u)\mathsf{d}B_{\eta }(u),\qquad u\in \lbrack
0,1],
\end{equation*}%
with $\sigma ^{2}(0)=\bar{\sigma}^{2}$, and where $(B_{\eta },B_{z})$ is a
bivariate standard Brownian motion.$\hfill \square $
\end{example}

\begin{remark}
In both examples, the process generating $\sigma _{t}^{2}$ depends on the
sample size $n$, so that $\left\{ \sigma _{nt}^{2}\right\} _{1\leq t\leq
n;n\geq 1}$ is actually a triangular array. We will not make this explicit
in the notation in this section.
\end{remark}

\begin{remark}
The initial condition $\sigma _{0}^{2}$ equals $\bar{\sigma}^{2}$ in both
examples; this value represents the unconditional variance in the GARCH
model, and $\log \bar{\sigma}^{2}$ is the (unconditionally) expected log-variance
in the stochastic volatility model, in both cases assuming $\kappa >0$.
However, this has only
been assumed for simplicity: the examples can be generalized to allow for an
arbitrary initial condition, fixed or random, as long as it satisfies $%
\sigma _{0}^{2}\overset{w}{\rightarrow }\sigma ^{2}(0)$ for some strictly
positive random variable $\sigma ^{2}(0)$, independent of $B_{\eta }$.
\end{remark}

\begin{remark}
The main difference between the examples is that in Example \ref{ex:SV}, the
volatility shocks $\{\eta _{t}\}_{t\geq 1}$ are independent of $%
\{z_{t}\}_{t\geq 1}$, whereas in Example \ref{ex:GARCH}, the variance is
driven by $\eta _{t}=(z_{t}^{2}-1)/\sqrt{2}$, which is fully determined by
(although uncorrelated with) $z_{t}$. In both examples, however, the limit
volatility process $\sigma (\cdot )$ is independent of the Brownian motion $%
B_{z}$ generated by $\{z_{t}\}$. If we think of $\varepsilon _{t}$ as the
deviation of a financial return from its conditional expectation, and $%
\{\sigma _{t}\}$ as its conditional volatility, then this rules out
so-called leverage effects, i.e., asymmetric effects of positive and
negative return shocks $\varepsilon _{t}$ on future volatility $\sigma
_{t+h} $, $h>0$. Although the results given in this section also apply to
processes with leverage, we will assume (asymptotic) independence in order
to establish bootstrap validity.
\end{remark}

\begin{remark}
In Example \ref{ex:SV}, the log-volatility follows a near-integrated
first-order autoregression, converging weakly to an Ornstein-Uhlenbeck [OU]
process. Similarly, in \cite{H95} $\{\sigma _{t}\}$ satisfies Assumption \ref%
{ass:SV} with $\sigma (\cdot )$ a (possibly nonlinear) transformation of an
OU process (or Brownian motion). Cases where the volatility is allowed to
jump at a countable number of times (while being constant between these jump
times) are also allowed by our assumption. For instance, let $\sigma
_{t}=\exp (\omega _{0}+\omega _{1}J_{t})$,$\text{ }J_{t}:=\sum_{i=1}^{t}%
\delta _{i}\eta _{i}$, $J_{0}=0$ a.s., where for all $t$, $\delta _{t}$ is a
Bernoulli random variable which equals one if and only if a volatility jump
occurs at time $t$. If the $\eta _{t}$'s (which denote the random jump
sizes) are i.i.d., independent of $\delta _{t}$'s, and if $P(\delta
_{t}=1)=\lambda n^{-1}$, then in $\mathscr{D}{}[0,1]$ (see
\citealp{G08}) $J_{n}(u):=J_{\lfloor nu\rfloor }$ converges weakly to the
compound Poisson process $C_{\lambda }(u):=\sum_{i=1}^{N(u)}\eta _{i}$,
where $N$ is a Poisson process in $\mathscr{D}{}[0,1]$ with intensity
parameter $\lambda $. As expected, the limiting volatility process is $%
\sigma (u)=\exp \left( \omega _{0}+\omega _{1}C_{\lambda }(u)\right) $, a
piecewise constant process with number of discontinuities given by $N(1)$.$%
\hfill \square $
\end{remark}

\subsection{Standard asymptotics under non-stationary stochastic volatility}

\label{sec standard asy}

Assumptions \ref{ass:mds} and \ref{ass:SV} allow to analyze the asymptotic
behavior of the functional $\left( M_{n},U_{n}\right) $, as is done in the
following Lemma.

\begin{lemma}
\label{lem:unc}Under Assumptions \ref{ass:mds} and \ref{ass:SV}, we have in $%
\mathscr{D}_{2}\left[ 0,1\right] $ as $n\rightarrow \infty $, 
\begin{equation}
\left( M_{n}(u),U_{n}(u)\right) \overset{w}{\rightarrow }\left(
M(u),V(u)\right) :=\left( \int_{0}^{u}\sigma (s)\mathsf{d}%
B_{z}(s),\int_{0}^{u}\sigma ^{2}(s)\mathsf{d}s\right) ,\quad u\in \lbrack
0,1],  \label{MVunc}
\end{equation}%
with $V(u)=\left\langle M\right\rangle (u)$. Furthermore, 
\begin{equation}
\sup_{u\in \lbrack 0,1]}|U_{n}(u)-V_{n}(u)|\overset{p}{\rightarrow }0.
\label{Vtilde}
\end{equation}
\end{lemma}

The implications for the testing problems in Section \ref{sec testing
problems} are given next.

\bigskip

\noindent \textsc{Example 1 (cont'd). }Consider the location model of
Example 1. A straightforward application of Lemma \ref{lem:unc} along with
the CMT yields that, under $\mathsf{H}_{0}$, $S_{n}=M_{n}(1)\overset{w}{%
\rightarrow }M(1)$, which corresponds to the mixed normal distribution $%
N(0,\int_{0}^{1}\sigma ^{2}(u)\mathsf{d}u)$ and hence is non-pivotal. For $%
T_{n}$ it holds that 
\begin{equation}
T_{n}=\frac{M_{n}(1)}{\sqrt{U_{n}(1)}}+o_{p}(1)\overset{w}{\rightarrow }%
\frac{M(1)}{\sqrt{V(1)}}=\frac{\int_{0}^{1}\sigma (u)\mathsf{d}B_{z}(u)}{%
\sqrt{\int_{0}^{1}\sigma ^{2}(u)\mathsf{d}u}}.  \label{eq asy tor T_n}
\end{equation}%
Notice that the limit distribution in (\ref{eq asy tor T_n}) is non-pivotal
in cases where $\sigma $ and $B_{z}$ are not stochastically independent. In
contrast, should independence hold, then (\ref{eq asy tor T_n}) corresponds
to a standard Gaussian distribution.

\bigskip

\noindent \textsc{Example 2 (cont'd)}. For the CUSUM test statistics of
Example 2 it holds that, again by Lemma \ref{lem:unc} and the CMT, that
under $\mathsf{H}_{0}$%
\begin{eqnarray*}
CS_{n}=\sup_{u\in \lbrack 0,1]}|M_{n}(u)-uM_{n}(1)| &\overset{w}{\rightarrow 
}&\sup_{u\in \lbrack 0,1]}|M(u)-uM(1)|, \\
CT_{n}=\frac{\sup_{u\in \lbrack 0,1]}|M_{n}(u)-uM_{n}(1)|}{\sqrt{%
U_{n}(1)-n^{-1}M_{n}^{2}(1)}} &\overset{w}{\rightarrow }&\frac{\sup_{u\in
\lbrack 0,1]}|M(u)-uM(1)|}{\sqrt{V(1)}}.
\end{eqnarray*}%
Both statistics have a non-pivotal asymptotic null distribution, even in
cases where the limit stochastic volatility process and the limit Brownian
motions are stochastically independent.

\bigskip

\noindent \textsc{Example 3 (cont'd)}. Finally, as shown in \cite{CT09}, for
the unit root testing problem the presence of non-stationary volatility
renders the null distribution of the Dickey-Fuller coefficient and $t$%
-statistics non-pivotal. More specifically, under the unit root null
hypothesis it holds that 
\begin{equation*}
R_{n}=\frac{\frac{1}{2}\left( M_{n}^{2}(1)-U_{n}(1)\right) }{%
\int_{0}^{1}M_{n}^{2}(u)\mathsf{d}u}\ \overset{w}{\rightarrow }\ \frac{\frac{%
1}{2}(M(1)^{2}-V(1))}{\int_{0}^{1}M^{2}(u)\mathsf{d}u}=\frac{\int_{0}^{1}M(u)%
\mathsf{d}M(u)}{\int_{0}^{1}M^{2}(u)\mathsf{d}u},
\end{equation*}%
and 
\begin{equation*}
W_{n}=\frac{\int_{0}^{1}M_{n}(u)\mathsf{d}M_{n}(u)}{\sqrt{%
\int_{0}^{1}M_{n}^{2}(u)\mathsf{d}u}}\frac{1}{\sqrt{U_{n}(1)+o_{p}(1)}}\ 
\overset{w}{\rightarrow }\ \frac{\int_{0}^{1}M(u)\mathsf{d}M(u)}{\sqrt{%
V(1)\int_{0}^{1}M^{2}(u)\mathsf{d}u}}
\end{equation*}%
as the sample size diverges.$\hfill \square $

\section{Bootstrap under non-stationary stochastic volatility}

\label{Sec invalidity of the BS}

Consider the standardized sample mean statistic $S_{n}$ for the location
model, see Example 1. Lemma \ref{lem:unc} implies that under the null
hypothesis $S_{n}\overset{w}{\rightarrow }M(1)=\int_{0}^{1}\sigma (u)\mathsf{%
d}B_{z}(u)$, see (\ref{MVunc}). The distribution of $M(1)$ depends on the
limit volatility process $\sigma $, implying that critical values cannot be
tabulated without providing a complete specification of this process. When
the Brownian motion $B_{z}$ and the limit volatility process $\sigma $ are
independent, then in the location model of Example 1 this problem can be
avoided by considering the studentized test statistic $T_{n}$, which has a
standard normal limit distribution under the null. However, in other testing
problems (such as those in Examples 2 and 3) it is generally not possible to
find such asymptotically pivotal statistics.

This motivates the development of bootstrap tests. Following much of the
literature (e.g.~\citealp{CT08,CT09}), we consider the wild bootstrap, which
replicates the volatility patterns in the original data. Let $w_{t}^{\ast }$
be an i.i.d.\ sequence with mean zero and variance  one\footnote{%
Further conditions on the moments of $w_{t}^{\ast }$ may be required in some
specific applications.}, independent of $\{\sigma _{t},z_{t}\}_{t\geq 1}$,
and define the bootstrap shocks as%
\begin{equation*}
\varepsilon _{t}^{\ast }=\varepsilon _{t}w_{t}^{\ast },\qquad t=1,2,\ldots
,n.
\end{equation*}%
Accordingly, we can define the bootstrap partial sum and the bootstrap
partial sum of squares, 
\begin{equation*}
\left( M_{n}^{\ast }(u),U_{n}^{\ast }(u)\right) =\left(
n^{-1/2}\textstyle\sum_{t=1}^{\lfloor nu\rfloor }\varepsilon _{t}^{\ast
},n^{-1}\textstyle\sum_{t=1}^{\lfloor nu\rfloor }\left( \varepsilon
_{t}^{\ast }\right) ^{2}\right) , \quad u\in \lbrack 0,1].
\end{equation*}%
These processes are the bootstrap analogs of the processes $M_{n}$ and $%
U_{n} $ of Section \ref{Sec Model}. Notice that this implementation of the
bootstrap assumes that $\varepsilon _{t}$ is observed under the null
hypothesis, which is the case in Examples 1 and 3 (location and unit root
test). In more general testing problems, including Example 2 (CUSUM test), $%
\varepsilon _{t}$ will be replaced by some residuals $\hat{\varepsilon}_{t}$
(either restricted by the null hypothesis or unrestricted).

\subsection{Failure of classic bootstrap validity}

\label{sec Anders is so boring today}

Classic validity of the bootstrap (usually denoted as `bootstrap
consistency') is usually understood as the convergence in probability (or
almost surely) of the conditional (on the original data) cdf of the
bootstrap statistic to the limit cdf of the original statistic. We show here
that, in the presence of stochastic volatility as in the previous section,
in general classic validity of the bootstrap fails. This is essentially
because the conditional cdf of the bootstrap statistic remains random in the
limit. In this section we discuss this fact and its implications on
bootstrap inference using, as the reference bootstrap algorithm, a wild
bootstrap scheme as is typically applied when the data are heteroskedastic.

Focusing again on the location statistic $S_{n}$, its bootstrap counterpart
is $S_{n}^{\ast }=n^{-1/2}\sum_{t=1}^{n}\varepsilon _{t}^{\ast
}=M_{n}^{\ast }(1)$ where $M_{n}^{\ast }$ is as previously defined. Define $%
P^{\ast }$ as the bootstrap measure conditional on the original data, $D_{n}$%
. The bootstrap (conditional) cdf is 
\begin{equation*}
F_{n}^{\ast }(x):=P^{\ast }(S_{n}^{\ast }\leq x)=P(S_{n}^{\ast }\leq x|D_{n})%
\text{.}
\end{equation*}

The classical condition for bootstrap validity is that, as $n\rightarrow
\infty $, $S_{n}^{\ast }\overset{w^{\ast }}{\rightarrow }_{p}S:=M(1)$. If
the limit cdf $F(x):=P(S\leq x)$ is continuous, then this weak convergence
in probability corresponds to the property 
\begin{equation}
\sup_{x\in \mathbb{R}}\left\vert F_{n}^{\ast }(x)-F(x)\right\vert \overset{p}%
{\rightarrow }0,  \label{wcp}
\end{equation}%
where $F$ denotes the cdf of the asymptotic distribution of $S_{n}$:%
\begin{equation*}
F(x):=P(N(0,V(1))\leq x)=\int \Phi (V(1)^{-1/2}x)\mathsf{d}P(V(1)),
\end{equation*}%
where $\Phi (\cdot )$ is the standard normal cdf. Because $F(x)$ is the
marginal cdf of $S=M(1)$, under independence of the processes $B_{z}$ and $%
\sigma $ it corresponds to the cdf of the mixed normal random variable $%
S=V(1)^{1/2}Z$, where $Z\sim N(0,1)$, independent of $V(1)$.

However, under Assumption \ref{ass:SV}, condition (\ref{wcp}) fails to hold,
which is seen as follows. Choosing $w_{t}^{\ast }\sim \mathrm{i.i.d.}~N(0,1)$
for convenience, it is seen that 
\begin{equation}
S_{n}^{\ast }|D_{n}\sim N(0,U_{n}(1))|U_{n}(1).
\label{eq distribution of S|D}
\end{equation}%
This follows because, conditional on the original data, 
\begin{equation*}
S_{n}^{\ast }=n^{-1/2}\textstyle\sum_{t=1}^{n}\varepsilon _{t}w_{t}^{\ast
}\sim N\left( 0,n^{-1}\textstyle\sum_{t=1}^{n}\varepsilon _{t}^{2} \right)\sim
N(0,U_{n}(1)).
\end{equation*}%
In terms of the conditional distribution $F_{n}^{\ast }$ of $S_{n}^{\ast }$
given the data $D_{n}$, (\ref{eq distribution of S|D}) corresponds to 
\begin{eqnarray*}
F_{n}^{\ast }(x)\overset{}{:=}P^{\ast }(S_{n}^{\ast }\leq x)
&=&P(N(0,U_{n}(1))\leq x|U_{n}(1)) \\
&=&P(N(0,1)\leq U_{n}(1)^{-1/2}x|U_{n}(1))=\Phi (U_{n}(1)^{-1/2}x).
\end{eqnarray*}%
Letting $n\rightarrow \infty $, the limit distribution of the bootstrap
statistic given the data follows from Lemma \ref{lem:unc} and the CMT.
Specifically, we have that 
\begin{equation}
F_{n}^{\ast }(x)\overset{w}{\rightarrow }\Phi (V(1)^{-1/2}x)
\label{eq convergence of F*}
\end{equation}%
for all $x\in \mathbb{R}$; eq.\ (\ref{eq convergence of F*}) implies that
the limit distribution of the bootstrap cdf is in fact \emph{random}. That
is, the bootstrap cdf $F_{n}^{\ast }$ does not converge in probability but
weakly, and the limiting cdf is random, as it depends on the random variable 
$V(1)$. Therefore, there is no reason to expect that the difference between
the random function $F_{n}^{\ast }$ and the non-random function $F$
converges in probability to $0$, as required for standard bootstrap
validity to apply.

The fact that the conditional cdf $F_{n}^{\ast }$ converges weakly (in $%
\mathscr{D}{}(\mathbb{R})$), rather than in probability, to a random cdf,
will be referred to as `weak convergence in distribution', and denoted as `$%
\overset{w}{\rightarrow }_{w}$'. More specifically, for sequences of random
variables $(Z_{n},Y_{n})$ and $(Z,Y)$ (possibly defined on different
probability spaces), the notation $Z_{n}|Y_{n}\overset{w}{\rightarrow }%
_{w}Z|Y$, when the conditional distribution of $Z|Y$ is diffuse
(non-atomic), means that 
\begin{equation}
F_{n}(\cdot |Y_{n}):=P(Z_{n}\leq \cdot |Y_{n})\overset{w}{\rightarrow }%
P(Z\leq \cdot |Y)=:F(\cdot |Y),\quad \text{in }\mathscr{D}{}(\mathbb{R}).
\label{eq weak convergence of the conditional cdf}
\end{equation}%
A more general definition of $Z_{n}|Y_{n}\overset{w}{\rightarrow }_{w}Z|Y$,
which includes non-diffuse conditional distributions and that is applicable
to the case where $(Z_{n},Y_{n})$ and $(Z,Y)$ are random elements of the
metric spaces $\mathcal{S}_{Z}\times \mathcal{S}_{Y_{n}}$ and $\mathcal{S}%
_{Z}\times \mathcal{S}_{Y}$, respectively (and hence to stochastic
processes), is that $E(g(Z_{n})|Y_{n})\overset{w}{\rightarrow }E(g(Z)|Y)$
for all bounded continuous functions $g:\mathcal{S}_{Z}\rightarrow \mathbb{R}
$, see \cite{CG20} and the references therein. We generalize this definition
to multivariate (joint) convergence in Section \ref{sec wcd and bootstrap
validity} below and employ it to prove bootstrap validity conditionally on
the volatility path.

When $Z_{n}$ represents a bootstrap statistic and the conditioning set $%
Y_{n} $ is the original data $D_{n}$, we use the notation `$\overset{w^{\ast
}}{\rightarrow }_{w}$'. Hence, eq.\ (\ref{eq convergence of F*}) corresponds
to the weak convergence in distribution%
\begin{equation*}
S_{n}^{\ast }\overset{w^{\ast }}{\rightarrow }_{w}N(0,V(1))|V(1)\text{.}
\end{equation*}%
Unless $V(1)$ is non-random (which is not the case under stochastic
volatility as considered here), this convergence shows that the limit
bootstrap measure is indeed a random measure. Hence, the bootstrap cannot be
valid in the usual sense of weak convergence in probability of $F_{n}^{\ast
} $ to $F$.

\subsection{Examples (continued)}

The result in Section \ref{sec Anders is so boring today} applies to the
other examples considered, except for the asymptotically pivotal statistic $%
T_{n}$.

\bigskip

\noindent \textsc{Example 1 (cont'd)}. Consider the location model example
and assume that the bootstrap data are generated as $y_{t}^{\ast
}=\varepsilon _{t}^{\ast }$, with $\varepsilon _{t}^{\ast }$ as defined
above. The bootstrap test statistics are $S_{n}^{\ast }:=\sqrt{n}\bar{%
\varepsilon}_{n}^{\ast }$ and $T_{n}^{\ast }:=\sqrt{n}\bar{\varepsilon}%
_{n}^{\ast }/s_{n}^{\ast }$, $s_{n}^{\ast
}=(n^{-1}\sum_{t=1}^{n}(\varepsilon _{t}^{\ast }-\bar{\varepsilon}_{n}^{\ast
})^{2})^{1/2}$. Using the argument discussed above, we have that $%
S_{n}^{\ast }\overset{w^{\ast }}{\rightarrow }_{w}M^{\ast }(1)|V(1)$, where $%
M^{\ast }(u):=\int_{0}^{u}\sigma (s)\mathsf{d}B_{z}^{\ast }(s)$ with $%
B_{z}^{\ast }$ a standard Brownian motion, stochastically independent of $%
\sigma $, and $V(1):=\int_{0}^{1}\sigma (u)^{2}\mathsf{d}u$. The conditional
distribution of $M^{\ast }(1)|V(1)$ is equal (a.s.) to the conditional
normal distribution $N(0,V(1))|V(1)$. In contrast, for $T_{n}^{\ast }$ it
holds that 
\begin{equation*}
T_{n}^{\ast }\overset{w^{\ast }}{\rightarrow }_{w}\left. \frac{M^{\ast }(1)}{%
\sqrt{V(1)}}\right\vert V(1),
\end{equation*}%
where the limit corresponds to a conditional $N(0,1)$ distribution
(independent of $V(1)$), following from the conditional $N(0,V(1))$
distribution of $M^{\ast}(1)\vert V(1)$. Since weak convergence of a
conditional distribution to a non-random cdf corresponds to weak convergence
in probability, in this special case $T_{n}^{\ast }\overset{w^{\ast }}{%
\rightarrow }_{p}N(0,1)$.

Two facts are worth stressing. First, in the above representations of the
limit conditional distribution of the bootstrap statistic, $\sigma $ and $%
B_{z}^{\ast }$ are independent, even if the original processes $\sigma $ and 
$B_{z}$ are not. This result stems from the assumption that the wild
bootstrap shocks $w_{t}^{\ast }$ are independent of the original data.
Second, if $\sigma $ and $B_{z}$ are stochastically independent, then the
conditional distributions of $M^{\ast }(1)|V(1)$ and $M(1)|V(1)$ are equal
(a.s). This distributional equality is crucial to determine validity of the
bootstrap.

\bigskip

\noindent \textsc{Example 2 (cont'd)}. For the bootstrap CUSUM statistics,
suppose that the bootstrap data are generated as $\varepsilon _{t}^{\ast }=%
\hat{\varepsilon}_{t}w_{t}^{\ast }$ where $\hat{\varepsilon}_{t}:=y_{t}-\bar{%
y}_{n}$, such that when the null hypothesis is true $\hat{\varepsilon}%
_{t}=\varepsilon _{t}-\bar{\varepsilon}_{n}$. The bootstrap statistics are
defined as 
\begin{equation*}
CS_{n}^{\ast }=\sup_{u\in \lbrack 0,1]}|M_{n}^{\ast }(u)-uM_{n}^{\ast
}(1)|,\quad CT_{n}^{\ast }=\frac{\sup_{u\in \lbrack 0,1]}|M_{n}^{\ast
}(u)-uM_{n}^{\ast }(1)|}{\sqrt{U_{n}^{\ast }(1)-n^{-1}M_{n}^{\ast }(1)^{2}}},
\end{equation*}%
with $M_{n}^{\ast }(u):=n^{-1/2}\sum_{t=1}^{\lfloor n\cdot \rfloor
}\varepsilon _{t}^{\ast }$ as above and $U_{n}^{\ast
}(1):=n^{-1}\sum_{t=1}^{n}(\varepsilon _{t}^{\ast })^{2}$. Under the null
hypothesis, 
\begin{equation*}
n^{-1/2}\sum_{t=1}^{\lfloor nu\rfloor }\varepsilon _{t}^{\ast
}=n^{-1/2}\sum_{t=1}^{\lfloor nu\rfloor }\hat{\varepsilon}_{t}w_{t}^{\ast
}=n^{-1/2}\sum_{t=1}^{\lfloor nu\rfloor }\varepsilon _{t}w_{t}^{\ast
}+O_{p}^{\ast }(n^{-1/2}),
\end{equation*}%
uniformly in $u\in \lbrack 0,1]$, and it holds that 
\begin{eqnarray*}
CS_{n}^{\ast } &\overset{w^{\ast }}{\rightarrow }_{w}& \sup_{u\in
\lbrack 0,1]} \left. |M^{\ast }(u)-uM^{\ast }(1)| \phantom{\tilde{I}} \hspace*{-1ex} \right\vert \sigma , \\
CT_{n} &\overset{w^{\ast }}{\rightarrow }_{w}&  \left. \frac{\sup_{u\in
\lbrack 0,1]}  |M^{\ast }(u)-uM^{\ast }(1)|}{\sqrt{V(1)}}\right\vert \sigma ,
\end{eqnarray*}%
where again $M^{\ast }(u):=\int_{0}^{u}\sigma (s)\mathsf{d}B_{z}^{\ast }(s)$
with $B_{z}^{\ast }$ a standard Brownian motion, stochastically independent
of $\sigma $, and $V(1):=\int_{0}^{1}\sigma (u)^{2}\mathsf{d}u$. Both
bootstrap statistics have a \emph{random }non-pivotal asymptotic null
distribution.

\bigskip

\noindent \textsc{Example 3 (cont'd)}. Finally, consider the unit root
example. To avoid the problems described in \cite{B91}, the bootstrap data
are generated with the unit root imposed, i.e.\ $y_{t}^{\ast }=y_{t-1}^{\ast
}+\varepsilon _{t}^{\ast }$, with $y_{0}^{\ast }=0$ and $\varepsilon
_{t}^{\ast }:=(\Delta y_{t})w_{t}^{\ast }$; see e.g.\ \cite{CT08}. Under the
null, clearly $\varepsilon _{t}^{\ast }:=\varepsilon _{t}w_{t}^{\ast }$. As
discussed earlier for the non-bootstrap case, we have that 
\begin{equation}
R_{n}^{\ast }=\frac{\frac{1}{2}\left( M_{n}^{\ast 2}(1)-U_{n}^{\ast
}(1)\right) }{\int_{0}^{1}M_{n}^{\ast 2}(u)\mathsf{d}u}\text{, }
\label{eq unit root R bootstrap statistic}
\end{equation}%
and, up to a negligible term, 
\begin{equation*}
W_{n}^{\ast }=\frac{\frac{1}{2}\left( M_{n}^{\ast 2}(1)-U_{n}^{\ast
}(1)\right) }{\sqrt{U_{n}^{\ast }(1)\int_{0}^{1}M_{n}^{\ast 2}(u)\mathsf{d}u}%
}.
\end{equation*}%
In this case it holds that 
\begin{equation}
R_{n}^{\ast }\overset{w^{\ast }}{\rightarrow }_{w}\left. \frac{%
\int_{0}^{1}M^{\ast }(u)\mathsf{d}M^{\ast }(u)}{\int_{0}^{1}M^{\ast 2}(u)%
\mathsf{d}u}\right\vert \sigma ,\quad W_{n}^{\ast }\overset{w^{\ast }}{%
\rightarrow }_{w}\left. \frac{\int_{0}^{1}M^{\ast }(u)\mathsf{d}M^{\ast }(u)%
}{\sqrt{V(1)\int_{0}^{1}M^{\ast 2}(u)\mathsf{d}u}}\right\vert \sigma ,
\label{asy distributions for the unit root bootstrap stats}
\end{equation}%
where $M^{\ast }(u):=\int_{0}^{u}\sigma (s)\mathsf{d}B_{z}^{\ast }(s)$ with $%
B_{z}^{\ast }$ a standard Brownian motion, stochastically independent of $%
\sigma $, and $V(u):=\int_{0}^{u}\sigma (s)^{2}\mathsf{d}s$. The asymptotic
distributions in (\ref{asy distributions for the unit root bootstrap stats})
are random, except in the special case where $\sigma $ is non-stochastic.$%
\hfill \square $

\bigskip

\citet{CG20} provide a number of other examples where the bootstrap validity
condition (\ref{wcp}) fails for any non-random cdf $F$, and develop an
alternative criterion for \emph{conditional} bootstrap validity. We will
apply this concept to the present situation and extend it to the analysis of
consistency of the bootstrap tests in the next section.

\section{Validity of the bootstrap}

\label{sec BS conditional validity}

Despite the fact that under non-stationary stochastic volatility the
bootstrap is unable to consistently estimate the limiting distribution of
the original statistic, it can still be valid, in the sense that it delivers
control over type one error probabilities as $n$ diverges. This can be seen
by focusing on the bootstrap \emph{p}-value. Taking the statistic $S_{n}$
and associated bootstrap analog $S_{n}^{\ast }$ to illustrate, the bootstrap 
\emph{p}-value is defined as%
\begin{equation*}
p_{n}^{\ast }:=P(S_{n}^{\ast }\leq S_{n}|D_{n})=F_{n}^{\ast }(x)|_{x=S_{n}},
\end{equation*}%
where $F_{n}^{\ast }(\cdot )$ is the cdf of $S_{n}^{\ast }$, conditional on
the data (this definition of the $p$-value assumes a left-tailed test, which
will be assumed below unless indicated otherwise). As in \citet{CG20}, we
say that the bootstrap based on $S_{n},S_{n}^{\ast }$ is valid conditionally
on the volatility process $\{\sigma _{t}\}_{t=1}^{n}$ if $p_{n}^{\ast }$ is
asymptotically $U(0,1)$ distributed conditionally on $\{\sigma
_{t}\}_{t=1}^{n}$, i.e.%
\begin{equation}
P(p_{n}^{\ast }\leq q|\{\sigma _{t}\}_{t=1}^{n})\overset{p}{\rightarrow }%
q,\quad q\in (0,1)\text{.}  \label{eq BS conditionally valid}
\end{equation}%
If this is the case, even if (as shown in the previous section) the limiting
conditional distribution of $S_{n}^{\ast }$ is random, the bootstrap test
can still be correctly sized in large samples. Moreover, proofs of validity
in the form of (\ref{eq BS conditionally valid}) also imply\footnote{%
For fixed $q\in (0,1)$, $Q_{n}:=P(p_{n}^{\ast }\leq q|\left\{ \sigma
_{t}\right\} _{t=1}^{n})$, is a random sequence in $[0,1]$, and hence
uniformly integrable. Result (\ref{eq BS conditionally valid}), i.e.\ $Q_{n}%
\overset{p}{\rightarrow }q$, together with uniform integrability of $Q_{n}$,
implies $L^{1}$ convergence and hence $P(p_{n}^{\ast }\leq
q)=E(Q_{n})\rightarrow q$ (see Kallenberg, 1997, Lemma 3.11).} that,
unconditionally, 
\begin{equation*}
P(p_{n}^{\ast }\leq q)\rightarrow q,\quad q\in (0,1).
\end{equation*}%
In the next subsections we discuss a set of sufficient conditions for (\ref%
{eq BS conditionally valid}) to hold. These are new in the literature on
bootstrapping conditional mean models when the volatility can be stochastic.
First, in Section \ref{sec wcd and bootstrap validity} we provide our
strategy to assess bootstrap validity. Our main results are given in Section %
\ref{sec main result}. Application to our examples are provided in Section %
\ref{sec examples-validity}. Finally, the behavior under the alternative is
analyzed and applied to our examples in Section \ref{sec behaviour under
alternative}.

\subsection{Weak convergence in distribution and bootstrap validity}

\label{sec wcd and bootstrap validity}

In this section, we summarize the approach developed by \citet{CG20},
applied here to establish conditional bootstrap validity in the presence of
non-stationary stochastic volatility. Before turning to bootstrap validity,
we introduce a multivariate version of the concept of weak convergence in
distribution that we anticipated in Section \ref{sec Anders is so boring
today}.

Precisely, let $Z_{n}:=(Z_{n}^{(1)},Z_{n}^{(2)})$ and $Z:=(Z^{(1)},Z^{(2)})$
be random elements of the (complete and separable)\ metric space $\mathcal{S}%
_{Z^{(1)}}\times \mathcal{S}_{Z^{(2)}}$, and $%
Y_{n}:=(Y_{n}^{(1)},Y_{n}^{(2)})$ and $Y:=(Y^{(1)},Y^{(2)})$ be random
elements of the (complete and separable)\ metric spaces $\mathcal{S}%
_{Y^{(1)}}^{\prime }\times \mathcal{S}_{Y^{(2)}}^{\prime }$ and $\mathcal{S}%
_{Y^{(1)}}\times \mathcal{S}_{Y^{(2)}}$, respectively. We say that 
\begin{equation}
(Z_{n}^{(1)}|Y_{n}^{(1)},Z_{n}^{(2)}|Y_{n}^{(2)})\overset{w}{\rightarrow }%
_{w}(Z^{(1)}|Y^{(1)},Z^{(2)}|Y^{(2)})  \label{eq joint weak conv}
\end{equation}%
jointly if, for all bounded continuous $g:\mathcal{S}_{Z^{(1)}}\rightarrow 
\mathbb{R}$ and $h:\mathcal{S}_{Z^{(2)}}\rightarrow \mathbb{R}$,%
\begin{equation*}
\left( E(g(Z_{n}^{(1)})|Y_{n}^{(1)}),E(h(Z_{n}^{(2)})|Y_{n}^{(2)})\right) 
\overset{w}{\rightarrow }\left(
E(g(Z^{(1)})|Y^{(1)}),E(h(Z^{(2)})|Y^{(2)})\right) \text{;}
\end{equation*}%
see \citet{K97} and \citet{CG20}. If $Z_{n}^{(i)},Z^{(i)}$, $i=1,2$ are
random variables and the conditional distributions $Z^{(1)}|Y^{(1)}$, $%
Z^{(2)}|Y^{(2)}$ are diffuse, then the above convergence is equivalent to 
\begin{eqnarray}
&&\left( P(Z_{n}^{(1)}\leq \cdot|Y_{n}^{(1)}),P(Z_{n}^{(2)}\leq
\cdot|Y_{n}^{(2)})\right)  \notag  \label{eq conv of cond cdf 2} \\
&&\hspace{2.5cm}\overset{w}{\rightarrow }\left( P(Z^{(1)}\leq
\cdot|Y^{(1)}),P(Z^{(2)}\leq \cdot|Y^{(2)})\right) \text{ in }\mathscr{D}{}(%
\mathbb{R}) \times \mathscr{D}{}(\mathbb{R}) . \qquad
\end{eqnarray}%
Finally, if $Z^{(1)}$ and $Z^{(2)}$ are the same random element (and
similary for $Y^{(1)}$ and $Y^{(2)}$), then we use the following alternative
notation for (\ref{eq joint weak conv}): 
\begin{equation*}
(Z_{n}^{(1)}|Y_{n}^{(1)},Z_{n}^{(2)}|Y_{n}^{(2)})\overset{w}{\rightarrow }%
_{w}(Z|Y,Z|Y)=:(1,1)Z|Y\text{, }
\end{equation*}%
and for $Z_{n}=(Z_{n}^{(1)},Z_{n}^{(2)})$ a bivariate random variable, (\ref%
{eq conv of cond cdf 2}) becomes%
\begin{equation*}
\left( P(Z_{n}^{(1)}\leq \cdot|Y_{n}^{(1)}),P(Z_{n}^{(2)}\leq
\cdot|Y_{n}^{(2)})\right) \overset{w}{\rightarrow }\left( \left( P(Z\leq
\cdot|Y\right) ,\left( P(Z\leq \cdot|Y\right) \right) \text{ in }%
\mathscr{D}{}(\mathbb{R}) \times \mathscr{D}{}(\mathbb{R}) .
\end{equation*}

Given this definition, we turn the attention to a statistic $\tau _{n}$
which is a function of the data $D_{n}$, which in our general set-up may be
represented by $(M_{n},U_{n})$; that is, $\tau _{n}=\tau (M_{n},U_{n})$. Its
bootstrap equivalent is $\tau _{n}^{\ast }=\tau (M_{n}^{\ast },U_{n}^{\ast
}) $, and we let $\tau =\tau (M,V)$, with $(M,V)$ denoting the weak limit of 
$\left( M_{n},U_{n}\right) $, see Section 2.3.

Recall that, with $\sigma _{n}\in \mathscr{D}{}[0,1]$ the volatility process
defined in (\ref{eq sigma_n(u) and B_z,n(u)}), conditioning on $\sigma _{n}$
is equivalent to conditioning on $\{\sigma _{t}\}_{t=1}^{n}$, and moreover
that $\sigma \in \mathscr{D}[0,1]$ is the weak limit of $\sigma _{n}$ by
Assumption \ref{ass:SV}. It then follows by \citet[Corollary 3.2]{CG20} that if the condition%
\begin{equation}
\left( \tau _{n}|\{\sigma _{t}\}_{t=1}^{n},\tau _{n}^{\ast }|D_{n}\right)
=\left( \tau _{n}|\sigma _{n},\tau _{n}^{\ast }|D_{n}\right) \overset{w}{%
\rightarrow }_{w}\left( \tau |\sigma ,\tau |\sigma \right) =\left(
1,1\right) \tau |\sigma  \label{jointww}
\end{equation}%
is satisfied, with the random cdf of $\tau |\sigma $ being sample-path
continuous, then 
\begin{equation*}
\sup_{x\in \mathbb{R}}\left\vert P(\tau _{n}\leq x|\{\sigma
_{t}\}_{t=1}^{n})-P(\tau _{n}^{\ast }\leq x|D_{n})\right\vert \overset{p}{%
\rightarrow }0.
\end{equation*}%
This means that the bootstrap consistently estimates the distribution of the
original statistic conditional on the volatility process, which in turn
implies that the bootstrap is conditionally valid, i.e., 
\begin{equation*}
P(p_{n}^{\ast }\leq q|\{\sigma _{t}\}_{t=1}^{n})\overset{p}{\rightarrow }q
\end{equation*}%
for all $q\in (0,1)$, where $p_{n}^{\ast }=P(\tau _{n}^{\ast }\leq \tau
_{n}|D_{n})$ is the bootstrap \emph{p}-value. The key condition to verify is
therefore the one given in (\ref{jointww}), along with continuity of the
limiting (random) cdf.

Because $\tau _{n}=\tau (M_{n},U_{n})$ with $(M_{n},U_{n})\in \mathscr{D}{}%
_{2}[0,1]$, proving (\ref{jointww}) involves proving conditional functional
limit theorems. It is known, see \citet{G94} and \citet{CP05}, that joint
weak convergence of e.g.\ $(\left( M_{n},U_{n}\right) ,\sigma _{n})$ is not
sufficient for conditional weak convergence. For example, \citet{G94} shows
that a sufficient condition is that $\left( M_{n},U_{n}\right) $ is
independent of $\sigma _{n}$, or that a change of measure can be found (with
weakly convergent Radon-Nikodym derivative) under which this independence
holds. These conditions do not seem to be directly applicable to the present
case.

\begin{remark}
An alternative approach to proving conditional limit theorems would be to
strengthen the joint weak convergence of $(\sigma _{n},B_{z,n})$ in
Assumption \ref{ass:SV} to $\mathcal{G}$-stable convergence, see \citet{HL15}%
, with $\mathcal{G}$ the $\sigma $-algebra generated by the limit volatility
process $\sigma (\cdot )$. This approach would require $\sigma _{n}$ and $%
\sigma $ to be defined on a common probability space, which is a natural
assumption in the analysis of bootstrapping realized volatility, see e.g.\ 
\cite{DGM13}; here $\sigma _{n}$ is a discretization of $\sigma $,
converging in probability to $\sigma $ by in-fill asymptotics. However, it
does not cover Example \ref{ex:GARCH}, where only weak convergence of $%
\sigma _{n}$ to $\sigma $ may be established, and the two processes are
defined on different probability spaces.

In contrast, our approach to proving (\ref{jointww}) involves Skorokhod's
representation theorem, and in particular the version of \citet{K97}, see
Corollary A.1 in the online appendix. This allows us to obtain limit
results \textquotedblleft as if\textquotedblright\ the conditioning element $%
\sigma _{n}$ converges almost surely to $\sigma $. By restricting the
dependence between $(M_{n},U_{n})$ and $\sigma _{n}$, we may then fix a
realization of the volatility process and prove an unconditional functional
limit theorem for each convergent sequence of realizations (except on a set
with measure zero). $\hfill \square $
\end{remark}

\subsection{Main results}

\label{sec main result}

Recall that the main assumption used to derive the limiting distribution of
the original statistic and the limiting (conditional) distribution of the
bootstrap statistic is that the errors form a mds with respect to the past
information set. This condition, however, is not sufficient for conditional
bootstrap validity, unless the volatility is deterministic or stationary. In
the presence of non-stationary stochastic volatility, further conditions are
required. A sufficient set of conditions is provided in the following
assumption.

\begin{assumption}
\label{ass:indep}Define $\mathcal{G}_{nt}:=\sigma\left( \left\{
z_{i}\right\} _{i=1}^{t},\left\{ \sigma_{i}\right\} _{i=1}^{n}\right) $ and
hence $\mathcal{G}_{n0}:=\sigma\left( \left\{ \sigma_{i}\right\}
_{i=1}^{n}\right) $, and define $\psi_{nt}^{2}:=E(z_{t}^{2}|\mathcal{G}%
_{n,t-1})$ and $v_{nt}:=z_{t}/\psi_{nt}$. Then:

\begin{enumerate}
\item[(a)] for all $n$, $\{v_{nt}\}_{t=1}^{n}$ is independent of $\mathcal{G}%
_{n0}$, and $\{\psi_{nt}\}_{t=1}^{n}$ is $\mathcal{G}_{n0}$-measurable.

\item[(b)] $\{z_{t},\mathcal{G}_{nt}\}_{1\leq t\leq n,n\geq 1}$ is a
martingale difference array (mda), satisfying for all $\epsilon >0$: 
\begin{equation}
n^{-1}\sum_{t=1}^{n}E\left (z_{t}^{2}\mathbb{I}_{\left\{ \left\vert
z_{t}\right\vert >\sqrt{n}\epsilon \right\} }|\mathcal{G}_{n0} \right )%
\overset{p}{\rightarrow }0.  \label{Lindeberg}
\end{equation}
\end{enumerate}
\end{assumption}

A few remarks are in order.

\begin{remark}
\label{rem:symmetry}If $\left\{ z_{t}\right\} _{t\geq 1}$ is independent of $%
\left\{ \sigma _{t}\right\} _{t\geq 1}$, as in Example \ref{ex:SV}, then
Assumption \ref{ass:indep} is trivially satisfied with $\psi _{nt}=1$ and $%
v_{nt}=z_{t}$ (the Lindeberg condition (\ref{Lindeberg}) is implicitly
assumed in Assumption \ref{ass:SV}, to guarantee $B_{z,n}\overset{w}{%
\rightarrow }B_{z}$). The dependence allowed by the assumption is needed to
cover situations such as the GARCH process in Example \ref{ex:GARCH}, where
for all $t<n$, 
\begin{equation*}
z_{t}^{2}=\frac{\sigma _{t+1}^{2}-\omega _{n}-\beta _{n}\sigma _{t}^{2}}{%
\alpha _{n}\sigma _{t}^{2}},
\end{equation*}%
which is known given $\left\{ \sigma _{t}\right\} _{t=1}^{n}$, such that $%
\psi _{nt}=\left\vert z_{t}\right\vert $ and hence $v_{nt}=z_{t}/\left\vert
z_{t}\right\vert =\mathrm{sgn}(z_{t})$, which because of symmetry of the
Gaussian distribution will indeed be independent of $\left\vert
z_{t}\right\vert $ and hence $\left\{ \sigma _{t}\right\} _{t=1}^{n}$.
\end{remark}

\begin{remark}
\label{rem:uncond}The mda assumption $E(z_{t}|\mathcal{G}_{n,t-1})=0$ rules
out leverage effects, such as implied by non-zero correlation between $z_{t}$
and the volatility shocks $\eta _{t}$ in the stochastic volatility model of
Example \ref{ex:SV}. It may be possible to weaken this assumption for the
results to follow, and allow for dependencies for finite $n$, as long as
they vanish asymptotically (such that $\sigma $ and $B_{z}$ are
independent). In the latter case it would be guaranteed that at least the
unconditional validity property $P(p_{n}^{\ast }\leq q)\rightarrow q$ for $%
q\in (0,1)$ holds for the bootstrap, by Theorem 3.1 of \citet{CG20}.
\end{remark}

\begin{remark}
\label{rem:indep}Part (a) of Assumption \ref{ass:indep} implies that we may
recover 
\begin{equation*}
M_{n}(\cdot )=n^{-1/2}\sum_{t=1}^{\lfloor n\cdot \rfloor }\sigma _{t}\psi
_{nt}v_{nt},\qquad U_{n}(\cdot )=n^{-1}\sum_{t=1}^{\lfloor n\cdot \rfloor
}\sigma _{t}^{2}\psi _{nt}^{2}v_{nt}^{2},
\end{equation*}%
and $V_{n}(\cdot )$ from the two independent sequences $\{\sigma
_{t}\}_{t=1}^{n}$ and $\{ v_{nt}\} _{t=1}^{n}$. This independence
facilitates the analysis of conditional distributions, as will be evident
from the proof of Theorem \ref{Thm: main}. $\hfill \square $
\end{remark}

The main result in this section is given in Theorem \ref{Thm: main} and
Corollary \ref{Cor: main} below.

\begin{theorem}
\label{Thm: main}Under Assumptions \ref{ass:mds}--\ref{ass:indep}, we have
as $n\rightarrow \infty $, 
\begin{equation*}
\left( 
\begin{array}{c}
\left( M_{n},U_{n}\right) |\ \sigma _{n} \\ 
\left( M_{n}^{\ast },U_{n}^{\ast }\right) |\ D_{n}%
\end{array}%
\right) \overset{w}{\rightarrow }_{w}\binom{1}{1}\left( M,V\right) \Big|%
\sigma \text{.}
\end{equation*}
\end{theorem}

The key result of Theorem \ref{Thm: main} is that the bootstrap processes $%
M_{n}^{\ast }$ and $U_{n}^{\ast }$, conditionally on the data, replicate in
the limit the distribution of the original processes $M_{n}$ and $U_{n}$,
conditionally on the volatility process $\sigma _{n}$ (or, equivalently, on $%
\{\sigma _{t}\}_{t=1}^{n}$). The implication of Theorem \ref{Thm: main} on
the behaviour of the bootstrap \emph{p}-values is provided in the following
corollary, which applies to a statistic $\tau _{n}=\tau (M_{n},U_{n})$
(which, under the null, converges weakly to $\tau =\tau (M,V)$) and its
bootstrap equivalent $\tau _{n}^{\ast }=\tau (M_{n}^{\ast },U_{n}^{\ast })$.

\begin{corollary}
\label{Cor: main}Under the conditions of Theorem \ref{Thm: main}, the
bootstrap is valid conditionally on $\{\sigma _{t}\}_{t=1}^{n}$, i.e.\ with $%
p_{n}^{\ast }:=P^{\ast }(\tau _{n}^{\ast }\leq \tau _{n})$,%
\begin{equation*}
p_{n}^{\ast }|\{\sigma _{t}\}_{t=1}^{n}\overset{w}{\rightarrow }_{w}U(0,1)%
\text{,}
\end{equation*}%
provided that the conditional distribution of $\tau =\tau (M,V)$ given $%
\sigma $ is sample-path continuous and the function $\tau $ is itself
continuous.
\end{corollary}

\subsection{Examples revisited}

\label{sec examples-validity}

In this section we check whether the conditions for bootstrap validity hold
for the examples. We assume throughout that Assumptions \ref{ass:mds} and %
\ref{ass:SV}, strengthened by \ref{ass:indep}, hold.

\bigskip

\noindent \textsc{Example 1 (cont'd)}. As earlier, the bootstrap statistics
are given by $S_{n}^{\ast }:=\sqrt{n}\bar{\varepsilon}_{n}^{\ast }$ and $%
T_{n}^{\ast }:=\sqrt{n}\bar{\varepsilon}_{n}^{\ast }/s_{n}^{\ast }$. Under
the null hypothesis, the original statistics are given by $S_{n}^{\ast }=%
\sqrt{n}\bar{\varepsilon}_{n}$ and $T_{n}^{\ast }=\sqrt{n}\bar{\varepsilon}%
_{n}/\hat{s}_{n}$. The original (bootstrap) statistics obtain as a
continuous transformation of $(M_{n},U_{n})$ (of $(M_{n}^{\ast },U_{n}^{\ast
})$). Hence by Theorem \ref{Thm: main} and a version of the CMT (see %
\citealp{CG20}, Theorem A.1) it holds that 
\begin{equation*}
\left( S_{n}|\{\sigma _{t}\}_{t=1}^{n},S_{n}^{\ast }|D_{n}\right) =\left(
S_{n}|\sigma _{n},S_{n}^{\ast }|D_{n}\right) \overset{w}{\rightarrow }%
_{w}(1,1)M(1)|\sigma \text{,}
\end{equation*}%
with $M(1)=\int_{0}^{1}\sigma (u)\mathsf{d}B_{z}(u)$. With $%
V(1)=\int_{0}^{1}\sigma (u)^{2}\mathsf{d}u$, the cdf of $M(1)|\sigma $ is
given by $\Phi (uV(1)^{-1/2})$, which is sample-path continuous with
probability 1. Hence, by Corollary \ref{Cor: main}, the bootstrap is valid
conditionally on the volatility path $\sigma $. For the studentized
statistic it holds that 
\begin{equation}
\left( T_{n}|\sigma _{n},T_{n}^{\ast }|D_{n}\right) \overset{w}{\rightarrow }%
_{w}(1,1)Z|\sigma \overset{a.s}{=}(1,1)Z\text{,}  \label{eq conv for T,T*}
\end{equation}%
where $Z\sim N(0,1)$, independent of $\sigma $; this implies that the
bootstrap is conditionally valid. Notice also that, see the discussion in
Section \ref{sec wcd and bootstrap validity}, (\ref{eq conv for T,T*}) and
continuity of the limiting Gaussian cdf $\Phi $ imply that $\sup_{x\in 
\mathbb{R}}|P^{\ast }(T_{n}^{\ast }\leq x)-\Phi (x)|=o_{p}\left( 1\right) $.
As $T_{n}\overset{w}{\rightarrow }Z$, the bootstrap test based on the
studentized statistics $T_{n},T_{n}^{\ast }$ is also valid in the classic
sense, i.e.\ the distribution of the bootstrap statistic $T_{n}^{\ast }$
conditional on the data consistenty estimates the (unconditional)\
distribution of the original statistic $T_{n}$. The same type of result does
not hold for the bootstrap based on $S_{n},S_{n}^{\ast }$; however,
bootstrap conditional validity is guaranteed by Corollary \ref{Cor: main}.

\bigskip

\noindent \textsc{Example 2 (cont'd)}. As for the previous example, since
the CUSUM (bootstrap) statistics are continuous transformations of $%
(M_{n},U_{n})$ (of $(M_{n}^{\ast },U_{n}^{\ast })$), from Theorem \ref{Thm:
main} and the CMT in \cite{CG20} we have that, for $\tau _{S}:=\sup_{u\in
\lbrack 0,1]}|M(u)-uM(1)|$ 
\begin{equation*}
\left( CS_{n}|\sigma _{n},CS_{n}^{\ast }|D_{n}\right) \overset{w}{%
\rightarrow }_{w}(1,1)\tau _{S}|\sigma \text{;}
\end{equation*}%
similarly, for $\tau _{T}:=V(1)^{-1/2}\sup_{u\in \lbrack 0,1]}|M(u)-uM(1)|,$ 
\begin{equation*}
\left( CT_{n}|\sigma _{n},CT_{n}^{\ast }|D_{n}\right) \overset{w}{%
\rightarrow }_{w}(1,1)\tau _{T}|\sigma \text{.}
\end{equation*}%
Both conditional asymptotic distributions are continuous with probability 1.
As discussed in \cite{A97}, this holds using the results in \cite{L82}
because the limiting random distribution corresponds (up to an almost surely
strictly positive term) to the supremum of a conditionally Gaussian process
with conditional covariance function which is nonsingular almost surely.
Hence, by Corollary \ref{Cor: main} the bootstrap is valid conditionally on
the volatility process $\sigma $.

\bigskip

\noindent \textsc{Example 3 (cont'd)}. Finally, in the unit root example we
have that, under the stated assumption and if the null hypothesis holds,
with $\tau _{R}:=(\int_{0}^{1}M^{2}(u)\mathsf{d}u)^{-1}\int_{0}^{1}M(u)%
\mathsf{d}M(u)$, 
\begin{equation*}
\left( R_{n}|\sigma _{n},R_{n}^{\ast }|D_{n}\right) \overset{w}{\rightarrow }%
_{w}(1,1)\tau _{R}|\sigma \text{.}
\end{equation*}%
Similarly, for the $t$ ratio test, with $\tau _{W}:=(\int_{0}^{1}M^{2}(u)%
\mathsf{d}u)^{-1/2}\int_{0}^{1}M(u)\mathsf{d}M(u)$, 
\begin{equation*}
\left( W_{n}|\sigma _{n},W_{n}^{\ast }|D_{n}\right) \overset{w}{\rightarrow }%
_{w}(1,1)\tau _{W}|\sigma \text{.}
\end{equation*}%
As proved in Lemma A.1 in the online appendix, the limiting conditional
cdfs have almost surely continuous sample paths, and hence by Corollary \ref%
{Cor: main}, the bootstrap is valid conditionally on $\sigma $.$\hfill
\square $

\subsection{Power considerations}

\label{sec behaviour under alternative}

We now briefly discuss the behavior of the bootstrap tests under the
alternative hypothesis when the stochastic volatility process induces
randomness of the limiting distribution of the bootstrap statistic. As
before, consider a left-sided test based on the statistic $\tau _{n}=\tau
(M_{n},U_{n})$ and its bootstrap equivalent $\tau _{n}^{\ast }=\tau
(M_{n}^{\ast },U_{n}^{\ast })$. Suppose that under the alternative the
original statistic diverges, say to $-\infty $, while the bootstrap
statistic satisfies 
\begin{equation}
\tau _{n}^{\ast }\overset{w^{\ast }}{\rightarrow }_{w}\tau ^{\ast} |\sigma
\label{eq conv for the BS}
\end{equation}%
for some random element $\tau ^{\ast} $. Then, the following lemma holds.

\begin{lemma}
\label{Lemma power}Suppose that (\ref{eq conv for the BS}) holds and that $%
\tau _{n}\overset{p}{\rightarrow }-\infty $ as $n\rightarrow \infty $. Then,
with $p_{n}^{\ast }:=P^{\ast }(\tau _{n}^{\ast }\leq \tau _{n})$, it holds
that $p_{n}^{\ast }\overset{p}{\rightarrow }0$.
\end{lemma}

Lemma \ref{Lemma power} shows that the fact that the limit distribution of
the bootstrap statistic is random and depends on the volatility path does
not affect the consistency of the bootstrap test. Essentially, weak convergence in
distribution of $\tau _{n}^{\ast }$ given the data implies that the
bootstrap statistic is $O_{p}^{\ast }(1) $, in probability. If the original
statistic diverges to $-\infty $, it then holds that the bootstrap test
rejects with probability converging to 1.\footnote{%
Notice that a consistent right-sided test can be obtained by focusing on the
bootstrap $p$-value $\tilde{p}_{n}^{\ast }:=1-p_{n}^{\ast }$.} We now apply
this result to the three leading examples.

\bigskip

\noindent \textsc{Example 1 (cont'd)}. Consider the location model example,
where the econometrician is interested in testing the simple null hypothesis 
$\theta =\bar{\theta}$ when $\bar{\theta}>\theta _{0}$, $\theta _{0}$ being
the true parameter value. A wild bootstrap with the null imposed generates
bootstrap data as $\varepsilon _{t}^{\ast }:=\hat{\varepsilon}%
_{t}w_{t}^{\ast }$ with $w_{t}^{\ast }$ i.i.d.\ $N(0,1)$ and $\hat{%
\varepsilon}_{t}:=y_{t}-\bar{\theta}=\varepsilon _{t}+\delta $, $\delta
:=\theta _{0}-\bar{\theta}<0$. It follows that, conditionally on the data, $%
M_{n}^{\ast }(\cdot ):=n^{-1/2}\sum_{t=1}^{\lfloor n\cdot \rfloor
}\varepsilon _{t}^{\ast }$ is a zero-mean Gaussian process with independent
increments and with conditional variance function $\hat{U}_{n}(\cdot )$,
where 
\begin{equation*}
\hat{U}_{n}(u):=n^{-1}\sum_{t=1}^{\lfloor nu\rfloor }\hat{\varepsilon}%
_{t}^{2}=U_{n}(u)+u\delta ^{2}+O_{p}(n^{-1/2})\overset{w}{\rightarrow }%
V(u)+u\delta ^{2},\quad \text{in }\mathscr{D}\text{.}
\end{equation*}%
Hence, under Assumptions \ref{ass:mds} and \ref{ass:SV}, we have, as $%
n\rightarrow \infty $, that\footnote{%
The proof reduces to the standard characterization of weak convergence to a
Gaussian process by means of a Skorokhod representation.} $M_{n}^{\ast }%
\overset{w^{\ast }}{\rightarrow }_{w}\tilde{M}^{\ast }|\sigma $, where $%
\tilde{M}^{\ast }(u):=\int_{0}^{u}\tilde{\sigma}(s)\mathsf{d}B_{z}^{\ast
}(s) $, with $\tilde{\sigma}(s):=(\sigma (s)^{2}+\delta ^{2})^{1/2}$. This
implies that $S_{n}^{\ast }\overset{w^{\ast }}{\rightarrow }_{w}\tilde{M}%
^{\ast }(1)|\sigma $. As $S_{n}\rightarrow -\infty $ as $n\rightarrow \infty 
$, the conditions of Lemma \ref{Lemma power} are satisfied and $p_{n}^{\ast }%
\overset{p}{\rightarrow }0$. Consistency of the test based on $T_{n}$
follows by standard arguments as $T_{n}^{\ast }\overset{w^{\ast }}{%
\rightarrow }_{p}N(0,1)$.

\bigskip

\noindent \textsc{Example 2 (cont'd)}. For the bootstrap CUSUM statistics,
consider the alternative $\theta _{t}=\theta _{1}+g(t/n)$, where $%
g:[0,1]\rightarrow \mathbb{R}$ is an arbitrary function satisfying $%
0<\int_{0}^{1}g^{2}(u)\mathsf{d}u<\infty $, see \cite{PK92}. The wild
bootstrap partial sum process is $M_{n}^{\ast
}(\cdot):=n^{-1/2}\sum_{t=1}^{\lfloor n\cdot\rfloor }\varepsilon
_{t}^{\ast }$, a zero-mean Gaussian process with independent increments and
with conditional variance function $\hat{U}_{n}(\cdot):=n^{-1}\sum%
_{t=1}^{\lfloor n\cdot\rfloor }\hat{\varepsilon}_{t}^{2}=U_{n}(%
\cdot)+G_{n}(\cdot)+O_{p}(n^{-1/2})$, with
\begin{equation*}
G_{n}(u):=n^{-1}\sum_{t=1}^{\lfloor nu\rfloor }\left(
g(t/n)-n^{-1}\sum_{t=1}^{n}g(t/n)\right) ^{2}\rightarrow \int_{0}^{u}\left(
g(s)-\int_{0}^{1}g(r)\mathsf{d}r\right) ^{2}\mathsf{d}s=:G(u),
\end{equation*}%
in $\mathscr{D}$. This implies that
\begin{equation*}
M_{n}^{\ast }(u)\overset{w^{\ast }}{\rightarrow }_{w}\tilde{M}^{\ast
}(u):= \left .  \int_{0}^{u} \tilde{\sigma}(s)\mathsf{d}B_{z}^{\ast }(s) \right | \sigma \text{,}
\quad \text{in } \mathscr{D} \text{,}
\end{equation*}%
with $\tilde{\sigma}(s):=(\sigma (s)^{2}+(g(s)-\int_{0}^{1}g(r)\mathsf{d}%
r)^2)^{1/2}$ and $B_{z}^{\ast }$ a standard Brownian motion, stochastically
independent of $\sigma $. The limiting distribution of the bootstrap
statistic $CS_{n}^{\ast }$ is then given by 
\begin{equation*}
CS_{n}^{\ast }\overset{w^{\ast }}{\rightarrow }_{w} \sup_{u\in \lbrack
0,1]}\left. |\tilde{M}^{\ast }(u)-u\tilde{M}^{\ast }(1)|\right | \sigma .
\end{equation*}%
In order to analyze the $CT_{n}^{\ast }$ statistic, notice that its
denominator satisfies, in probability, 
\begin{equation*}
U_{n}^{\ast }(1)-n^{-1}M_{n}^{\ast }(1)^{2}=U_{n}^{\ast }(1)+O_{p}^{\ast
}(n^{-1})=\hat{U}_{n}(1)+O_{p}^{\ast }(n^{-1/2})\overset{w}{\rightarrow }%
V(1)+G(1),
\end{equation*}%
which implies that%
\begin{equation*}
CT_{n}^{\ast }\overset{w^{\ast }}{\rightarrow }_{w}\left. \frac{\sup_{u\in
\lbrack 0,1]}|\tilde{M}^{\ast }(u)-u\tilde{M}^{\ast }(1)|}{\sqrt{V(1)+G(1)}}%
\right\vert \sigma \text{.}
\end{equation*}%
As both $S_{n}$ and $T_{n}$ diverge under the alternative considered, Lemma %
\ref{Lemma power} applies and for both tests $p_{n}^{\ast }\overset{p}{%
\rightarrow }0$.

\bigskip

\noindent \textsc{Example 3 (cont'd)}. Consider the unit root example with
wild bootstrap shocks generated with the null hypothesis, i.e. $\varepsilon
_{t}^{\ast }:=(\Delta y_{t})w_{t}^{\ast }$. The bootstrap $R_{n}^{\ast }$
statistic is given as in (\ref{eq unit root R bootstrap statistic}) with $%
M_{n}^{\ast }(u):=n^{-1/2}\sum_{t=1}^{\lfloor nu\rfloor }(\Delta
y_{t})w_{t}^{\ast }$ and $U_{n}^{\ast }(u):=n^{-1/2}\sum_{t=1}^{\lfloor
nu\rfloor }(\Delta y_{t})^{2}w_{t}^{\ast }{}^{2}$. Conditionally on the
data, $M_{n}^{\ast }$ is a zero-mean Gaussian process with independent
increments and conditional variance function $\hat{U}_{n}(\cdot):=n^{-1}%
\sum_{t=1}^{\lfloor n\cdot\rfloor }(\Delta y_{t})^{2}$. Under the
alternative that $y_{t}=(1+\theta )y_{t-1}+\varepsilon _{t}$ with $\theta
\in (-2,0)$, $\Delta y_{t}$ can be written as the linear process with
exponentially decaying coefficients $\Delta y_{t}=\sum_{i=0}^{t-1}\psi
_{i}\varepsilon _{t-i}$ with $\psi _{0}=0$ and $\psi _{i}=\theta (1+\theta
)^{i-1}$, $i=1,2,\ldots $. Hence, by standard decompositions for squared
stationary autoregressions it holds that (the proof is reported in the
online appendix)%
\begin{equation}
\hat{U}_{n}(u)=\overline{\psi }^{2}U_{n}(u)+o_{p}(1),\qquad \overline{\psi }%
^{2}:=\sum_{i=0}^{\infty }\psi _{i}^{2},  \label{eq in example 3 part 3}
\end{equation}%
where the $o_{p}(1)$ term is uniform in $\cdot \in \lbrack 0,1]$, which
implies that $\hat{U}_{n}\overset{w}{\rightarrow }\overline{\psi }^{2} V$.
Hence, $M_{n}^{\ast }\overset{w^{\ast }}{\rightarrow }_{w} \overline{%
\psi } M^{\ast }|\sigma $. Finally, using the fact that $U_{n}^{\ast }=%
\overline{\psi }^{2}\hat{U}_{n}+o_{p}^{\ast }(1)$, in probability, where $%
\hat{U}_{n}\overset{w}{\rightarrow }V$, it holds that%
\begin{equation*}
R_{n}^{\ast }\overset{w^{\ast }}{\rightarrow }_{w}\left. \frac{\frac{1}{2}(%
\overline{\psi }^{2}M^{\ast 2}(1)-\overline{\psi }^{2}V(1))}{\overline{\psi }%
^{2}\int_{0}^{1}M^{\ast 2}(u)\mathsf{d}u}\right\vert \sigma =\left. \frac{%
\frac{1}{2}(M^{\ast 2}(1)-V(1))}{\overline{\psi }^{2}\int_{0}^{1}M^{\ast
2}(u)\mathsf{d}u}\right\vert \sigma \text{.}
\end{equation*}%
Hence, under the alternative the bootstrap replicates the null distribution
of the original statistic conditional on the volatility process and
consistency of the bootstrap test follows from Lemma \ref{Lemma power}. An
identical result holds for the $t$-ratio test based on $W_{n}$.$\hfill
\square $

\bigskip

\begin{remark}
\label{Rem when anders was sleeping}While in this section we focused on
asymptotic power against fixed alternatives, it is possible to extend our
analysis to cover power against local alternatives. To illustrate, consider
the test based on $S_{n}$ for the hypothesis $\mathsf{H}_{0}:\theta =\bar{%
\theta}$ in the location model $y_{t}=\theta +\varepsilon _{t}$ (Example 1).
Under a sequence of local alternatives of the form $\mathsf{H}_{n}:\theta
_{n}=\bar{\theta}+\delta _{n}$ with $\delta _{n}=n^{-1/2}c$, it is
straightforward to show that $S_{n}=c+M_{n}(1)$, which converges weakly to $%
c+M(1)$ under Assumptions 1--2. For the bootstrap statistic, the results
obtained above with $\delta _{n}\rightarrow 0$ imply $S_{n}^{\ast }\overset{%
w^{\ast }}{\rightarrow }_{w}M(1)^{\ast }|V(1)=V(1)^{1/2}Z^{\ast }|V(1)$,
with $Z^{\ast }$ being $N(0,1)$ (independent of $V(1)$); hence, the
bootstrap distribution under the local alternative is the same as under the
null. Suppose now that Assumption 3 holds; then, by Theorem \ref{Thm: main}, 
\begin{equation*}
\left( S_{n}|\sigma _{n},S_{n}^{\ast }|D_{n}\right) \overset{w}{\rightarrow }%
_{w}\left( c+M(1)|\sigma ,M(1)|\sigma \right) .
\end{equation*}%
Hence the limiting cdf of $S_{n}|\{\sigma _{t}\}_{t=1}^{n}$ is given by $%
F_{c}(x)=\Phi ((x-c)V(1)^{-1/2})$, which is continuous with probability 1,
while the bootstrap cdf $F_{n}^{\ast }(x)$ converges weakly to $F^{\ast
}(x)=\Phi (xV(1)^{-1/2})$. Then, by application of Theorem 3.3 in \cite{CG20}
it holds that the power of the bootstrap test at the $100\alpha \%$ nominal
level, conditionally on the volatility process, is given by 
\begin{eqnarray*}
P(p_{n}^{\ast }\leq \alpha |\sigma _{n})=P(F_{n}^{\ast }(S_{n})\leq \alpha
|\sigma _{n}) &\overset{w}{\rightarrow }&F_{c}(F^{\ast -1}(\alpha )) \\
&=&\Phi ((V(1)^{1/2}\Phi ^{-1}(\alpha )-c)V(1)^{-1/2}) \\
&=&\Phi (\Phi ^{-1}(\alpha )-cV(1)^{-1/2}).
\end{eqnarray*}%
By construction, the local power function depends on $c$; we observe that it
also depends on $V(1)$, and hence is random in the limit. In more general
testing problems, the conditional local power function will depend on the
entire volatility process. In the next section, we illustrate, by Monte
Carlo simulations, the dependence on $c$ as well as on (the limit of) $%
\sigma _{n}$. $\hfill \square $
\end{remark}

\section{Numerical results}

\label{Sec MC}In this section we analyze finite sample size and power
properties of bootstrap tests under non-stationary stochastic volatility
using Monte Carlo simulations. To study the behavior of the tests from
Examples 1--3 under the null hypothesis, we report the Monte Carlo
(empirical) cdfs of bootstrap $p$-values, both unconditionally over all
Monte Carlo replications and conditionally on specific simulated volatility
paths. Following \cite{CG20}, we report the results in the form of fan
charts of the conditional cdfs, displayed together with the unconditional
cdf and the theoretical cdf of the $U(0,1)$ distribution. Similarly, we
display conditional power curves of the tests under local alternatives in
fan charts.

In all experiments, we draw observations $\{\varepsilon _{t}\}_{t=1}^{n}$
from the GARCH(1,1) process from Example V.2, with 
\begin{equation*}
\omega _{n}=1-\alpha _{n}-\beta _{n}=n^{-1}\kappa ,\qquad \alpha
_{n}=(2n)^{-1/2}\sigma _{\eta },
\end{equation*}%
corresponding to a limit process $\sigma ^{2}(u)$ with unit unconditional
variance $\bar{\sigma}^{2}=1$, mean-reversion parameter $\kappa $, and
volatility-of-volatility parameter $\sigma _{\eta }$. The conditional
variance sequence is initialized at the unconditional variance, i.e., $%
\sigma _{1}^{2}=1$. We set $\kappa =5$ and $\sigma _{\eta }=\sqrt{10}$,
corresponding to a rather persistent volatility process with a reasonable
amount of short-run variability of the volatility, which we know from
earlier simulation studies to lead to substantial size distortions in tests
based on standard (constant-volatility) asymptotic critical values. We expect
similar results from stochastic volatility processes (Example V.1) with the
same type of persistence and volatility-of-volatility properties. We report
results for two sample sizes, $n\in \{100,500\}$. The standardized errors $%
z_{t}$ are drawn from three different distributions, discussed below.

For each distribution and sample size, we first simulate $100$ different
realizations of the volatility path $\{\sigma _{t}\}_{t=1}^{n}$. For each of
these paths, we draw $50,000$ replications from the conditional distribution
of $\{\varepsilon _{t}=\sigma _{t}z_{t}\}_{t=1}^{n}$ given $\{\sigma
_{t}\}_{t=1}^{n}$. As discussed in Remark \ref{rem:symmetry}, this is
equivalent to drawing $v_{nt}=\mathrm{sgn}(z_{t})$ conditional on $\psi
_{nt}=|z_{t}|$ for $t=1,\ldots ,n-1$, and drawing $z_{n}$ from its
unconditional distribution (independent of $\{\sigma _{t}\}_{t=1}^{n}$). For
each choice of the distribution of $\{z_{t}\}_{t=1}^{n}$, we can check the
conditions of Assumption \ref{ass:indep} for conditional validity of the
bootstrap.

The first data-generating process, labelled DGP 1, is defined by $z_{t}\sim
N(0,1)$. In that case the conditional distribution of the signs $v_{nt}$ is
discrete uniform over $\{-1,1\}$, independent of $|z_{t}|$. This in turn
implies that the mda condition of Assumption \ref{ass:indep} is satisfied
(as well as the independence, measurability and Lindeberg conditions), such
that the bootstrap is conditionally valid.

\begin{figure}[t]
\centerline{\includegraphics[width=0.95\textwidth]{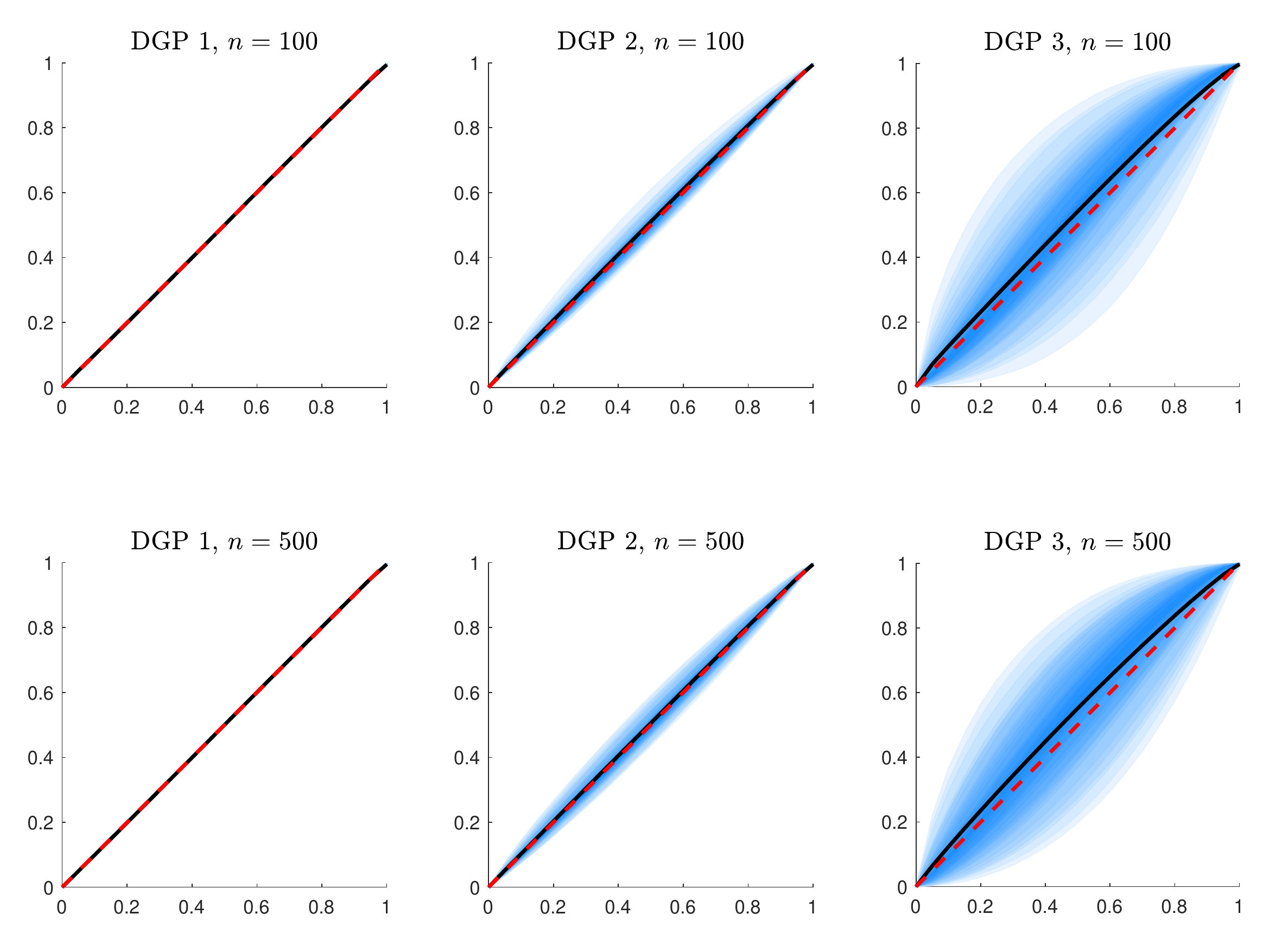}}
\caption{Monte Carlo conditional cdfs of bootstrap $p$-values of $T_{n}$}
\end{figure}

In DGP 2, $z_{t}$ is drawn from the following mixed normal density 
\begin{equation}
f(z)=\tfrac{1}{3}\phi (z;\mu _{1},\sigma _{1})+\tfrac{2}{3}\phi (z;\mu
_{2},\sigma _{2}),  \label{eq:MN}
\end{equation}%
where $\phi (z;\mu ,\sigma )$ is the pdf of the $N(\mu ,\sigma ^{2})$
distribution, and where $\mu _{1}=-2a,\sigma _{1}=a,\mu _{2}=a,\sigma _{2}=a%
\sqrt{2}$, with $a=\sqrt{3/11}$. This distribution was constructed by \cite%
{M00} to be asymmetric but with skewness $0$ (and with mean zero and unit
variance). Because $v_{nt}=\mathrm{sgn}(z_{t})$ in this case has a
conditional distribution depending on $\psi _{nt}=|z_{t}|$, with $%
P(v_{nt}=1|\psi _{nt})=f(\psi _{nt})/(f(\psi _{nt})+f(-\psi _{nt}))\neq 
\frac{1}{2}$, it follows that Assumption \ref{ass:indep} is violated, and
conditional validity of the bootstrap is not guaranteed. On the other hand,
the zero skewness implies that the limit result of Example V.2 still
applies, with $B_{z}$ independent of $B_{\eta }$ and hence $\sigma $. As
conjectured in Remark \ref{rem:uncond}, we may expect unconditional
bootstrap validity in this case.

\begin{figure}[t]
\centerline{\includegraphics[width=0.95\textwidth]{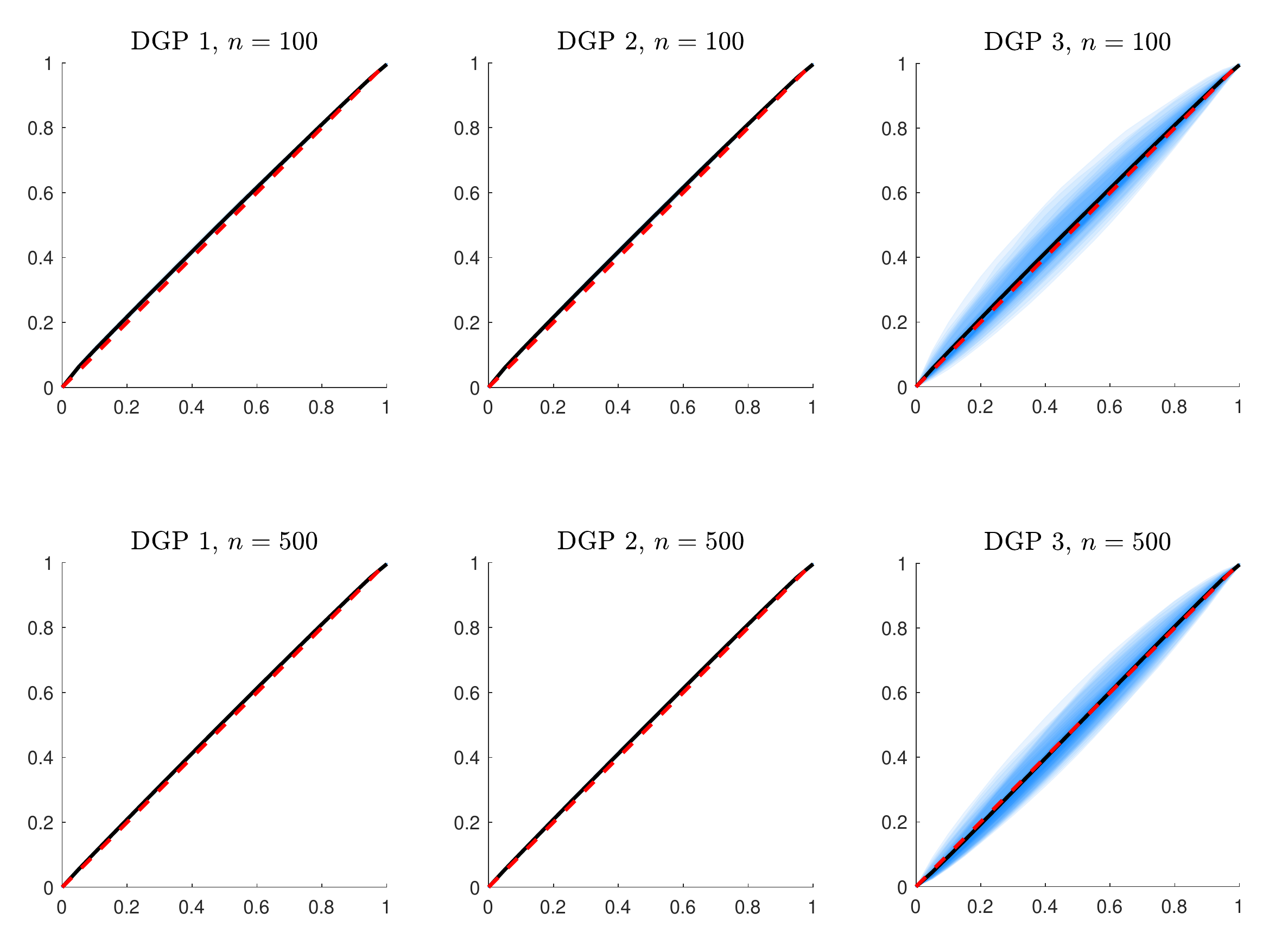}}
\caption{Monte Carlo conditional cdfs of bootstrap $p$-values of $CT_{n}$}
\end{figure}

In DGP 3, $z_{t}$ is drawn from another version of (\ref{eq:MN}), but now
with $\mu _{1}=-2b,\sigma _{1}=b\sqrt{2},\mu _{2}=b,\sigma _{2}=b$, with $b=%
\sqrt{3/10}$. This is a distribution with mean zero, unit variance and
negative skewness, so that $B_{z}$ and $B_{\eta }$ in Example V.2 have a
negative correlation, corresponding to long-run leverage effects. This implies that
the wild bootstrap is invalid in this case, both conditionally and
unconditionally.

Figures 1--3 display the results for the behavior of bootstrap $p$-values
(based on $199$ bootstrap replications) under the null hypothesis, for the
studentized tests based on $T_{n}$, $CT_{n}$ and $W_{n}$, respectively.
Unreported results for the other three test statistics $S_{n}$, $CS_{n}$ and 
$R_{n}$ are very similar to the results for the corresponding studentized
tests.

\begin{figure}[t]
\centerline{\includegraphics[width=0.95\textwidth]{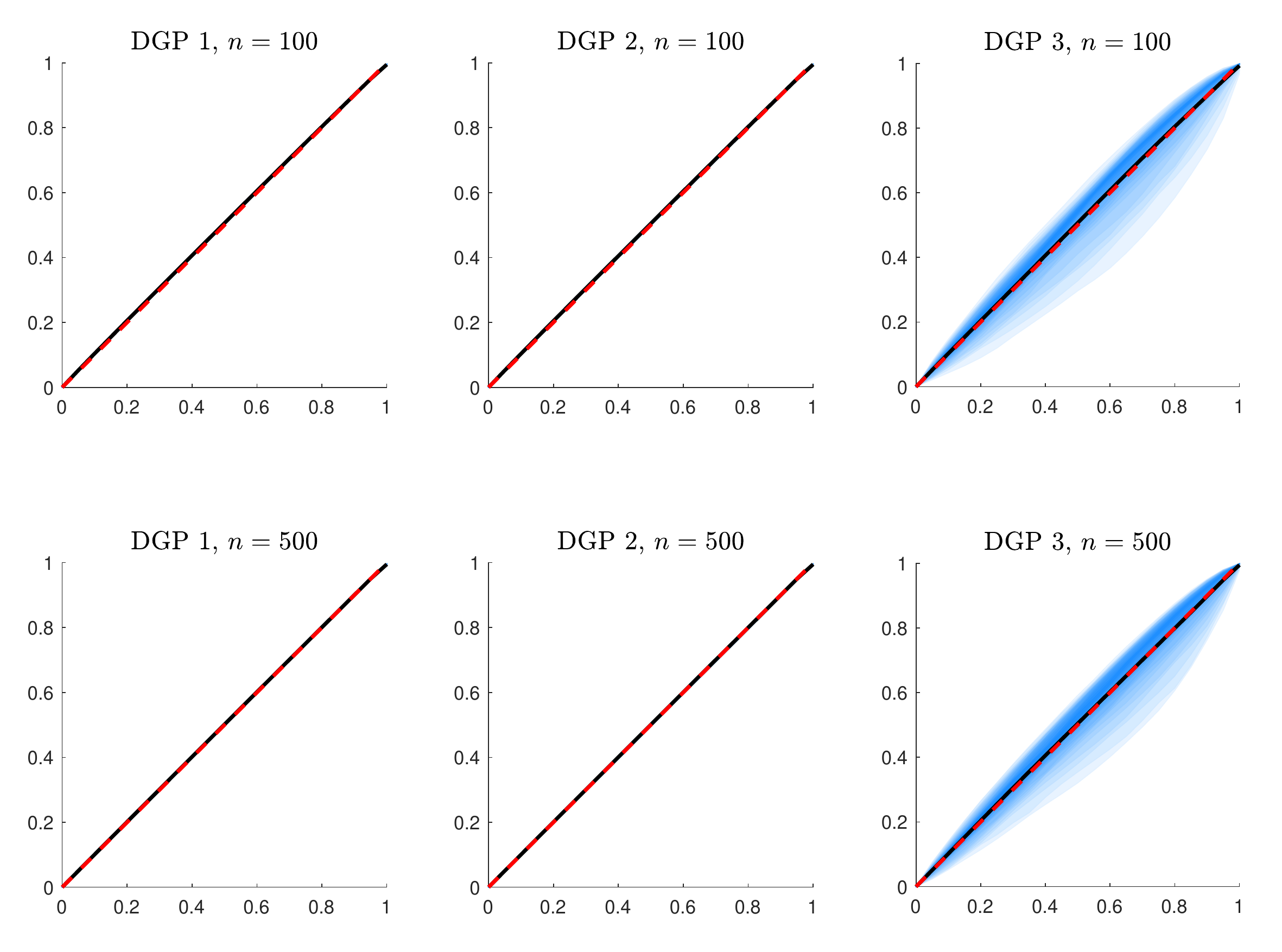}}
\caption{Monte Carlo conditional cdfs of bootstrap $p$-values of $W_{n}$}
\end{figure}

From Figure 1, we observe that when the standardized errors $z_{t}$ are
standard normal (DGP 1, left panels), then the conditional distribution of
the bootstrap $p$-values is very close to uniform, and appears to be
independent of $\sigma $ for both sample sizes considered. Thus the
theoretical conditional validity of the bootstrap in this case is clearly
reflected in finite-sample behavior. When the distribution of $z_{t}$ is
asymmetric with zero skewness (DGP 2, centre panels), then the bootstrap
appears to be valid on average (indicated by the solid line almost
coinciding with the $U(0,1)$ cdf, especially for $n=500$), but the
conditional cdfs of bootstrap $p$-values do depend on the volatility path
and deviate from the uniform cdf, illustrating the conjectured violation of
conditional bootstrap validity. Finally, for DGP 3 (right panels, skewed $%
z_{t}$), we observe more extreme dependence of bootstrap $p$-values on the
volatility path. In this case the bootstrap does not appear to be valid on
average either, as predicted by the dependence between $B_{z}$ and $\sigma $
implied by this DGP, which is not replicated by the wild bootstrap.

Figure 2 displays the results for the studentized CUSUM test based on $CT_n$%
. For this test, the finite-sample size distortion (indicated by the
difference between the solid and dashed line) is more pronounced than for
the location test, in particular for the smaller sample size ($n=100$).
Unreported additional simulations show that these size distortions are even
stronger for the test based on $CS_n$. The results improve when the sample
size increases, and it should be noted that the rejection frequencies at the 
$5\%$ significance level are still fairly close to $0.05$; the deviations
are larger at the centre of the distribution. For this test, the dependence
of the conditional cdf of $p$-values on $\sigma$ is much weaker than for the
location test. For DGP 2, we do not observe any deviation of conditional
cdfs from their average; in case of DGP 3, there is a clear violation of
conditional bootstrap validity, but the deviations are less pronounced than
for $T_n$.

The results for the unit root test based on $W_{n}$ are given in Figure 3.
For DGP 1 and 2, the size distortions appear to be negligible for both
sample sizes. Similarly to the CUSUM test, the dependence of bootstrap $p$%
-values on the volatility path for DGP 1 and 2 appears to be very weak. On
the other hand, for DGP 3 we find this dependence to be clearly present,
illustrating again a violation of conditional bootstrap validity. On
average, the bootstrap appears to be valid even for DGP 3, although
theoretically we would not expect this to be the case because of the
dependence between $B_{z}$ and $\sigma $. Unreported simulations have shown
that in case of stronger leverage effects (i.e., a stronger correlation
between $B_{z}$ and $B_{\eta }$), the unconditional cdf of bootstrap $p$%
-values does differ from the $U(0,1)$ cdf, as predicted by the theoretical
results.

\begin{figure}[t]
\centerline{\includegraphics[width=0.95\textwidth]{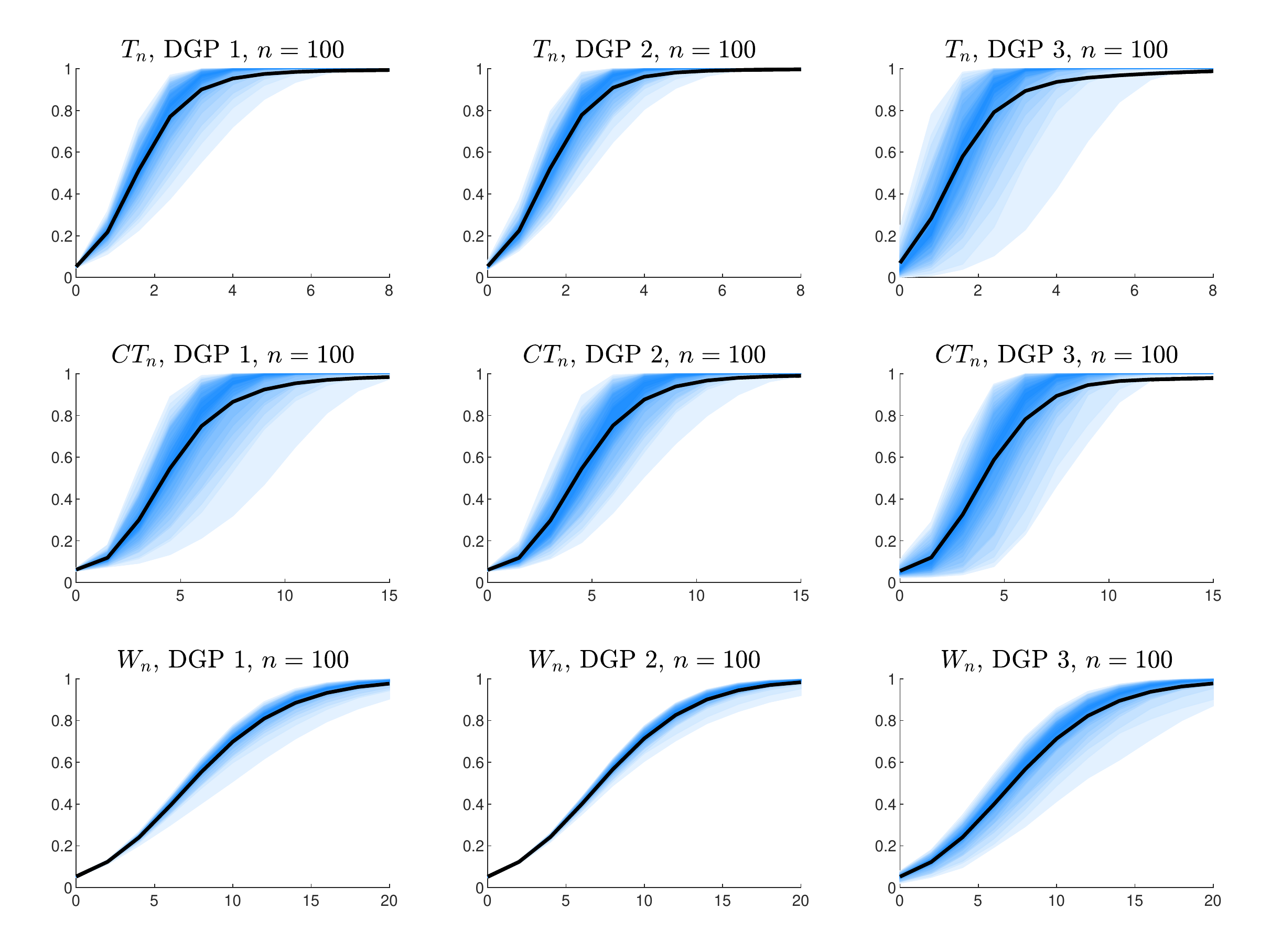}}
\caption{Monte Carlo conditional rejection frequencies of bootstrap tests}
\end{figure}

Next, we investigate the local power of the bootstrap tests, again
conditional on the same realizations of the volatility path as considered
for the size of the tests. As in Remark \ref{Rem when anders was sleeping},
for the location tests we evaluate the rejection frequency of the test for $%
\mathsf{H}_{0}:\theta =0$ against local alternatives $\theta _{n}=-n^{-1/2}c$%
, with $c\in \lbrack 0,8]$. For the CUSUM tests, the local alternative is a
break in the mean of the series, at $t=n/2$, from $\theta _{n,t}=0$ to $%
\theta _{n,t}=n^{-1/2}c$, with $c\in \lbrack 0,15]$. For the unit root
tests, we consider local alternatives $\theta _{n}=-n^{-1}c$, with $c\in
\lbrack 0,20]$. We provide results for the tests based on the studentized
statistics $T_{n}$, $CT_{n}$ and $W_{n}$, and for the sample size $n=100$.

Figure 4 displays the rejection frequencies of the bootstrap tests, based on 
$10,000$ replications of the test for each volatility path, plotted against $%
c$. We observe that the conditional rejection probabilities under the
alternative hypothesis depend on both the non-centrality parameter $c$ and
the volatility process $\sigma $, for each test and DGP 1--3. While the
dependence on the volatility is as expected for DGP 2 and 3, we note that
for DGP 1, where the rejections probabilities under the null hypothesis are
conditionally independent of the volatility process, the power of the tests
clearly depends on the volatility (as discussed in Remark \ref{Rem when
anders was sleeping}).

To gain some insight in the sensitivity of the simulation results to the
chosen parameter values, we have repeated the Monte Carlo experiments with $%
\kappa =0$, implying an integrated GARCH process ($\alpha _{n}+\beta _{n}=1$%
) with infinite unconditional variance. As in the original simulations, the
initial condition is $\sigma _{1}^{2}=1$, the volatility of volatility is $%
\sigma _{\eta }=\sqrt{10}$, and the same three DGPs for $z_{t}$ have been
used. The corresponding figures (available from the authors upon request)
show qualitatively very similar results as those displayed in Figures 1--4.
The increased persistence and variation in the realizations of $\{\sigma
_{t}^{2}\}_{t=1}^{n}$ leads to a bigger bias in DGP 3 for the $W_{n}$ test
statistic, and a bit more variation in the conditional distributions for DGP
3 for the $W_{n}$ and $CT_{n}$ test statistics. Furthermore, we find more
variation in conditional power functions and a slightly higher average
(unconditional) power.

\section{Conclusions}

\label{Sec conclusions}

In this paper we have analyzed the properties of wild bootstrap inference in
time series models for the conditional mean under non-stationary stochastic
volatility. In our setting, we do not make any specific assumption on the
volatility process, rather than assuming that it admits a weak limit in $%
\mathscr{D}{}[0,1]$. The additional advantage of this semi-parametric nature
of our analysis is that we do not need any assumptions on the existence of
higher moments, or on the initial condition of the conditional variance
process. On the other hand, if the parametric form of the volatility process
was known to the econometrician, then other model-based estimators and tests
(such as \cite{S99}, for the unit root testing problem under GARCH(1,1)
errors) and related bootstrap methods could be implemented. These
approaches, although particularly interesting and likely to deliver power
improvements over our wild bootstrap methods, are beyond the scope of this
paper.

A central ingredient in our analysis is that the bootstrap distribution (the
distribution of a bootstrap statistic conditional on the data) has a random
limit. In our case, this random limit can be characterized as the
conditional distribution of the limiting statistic conditional on a
continuous-time volatility process (which is itself the weak limit of the
finite-sample volatility process). The random nature of the limit of
bootstrap distributions is not new: a well-known example in the unit-root
literature is \cite{B91}, and \cite{SBW10} provide another example from
non-parametric statistics involving cube-root asymptotics. \cite{CG20}
analyze a number of other examples of this phenomenon. In the examples
encountered so far, this random limit distribution can always be expressed
as a conditional distribution, but its specific form is determined by a
combination of model assumptions, statistical methods and bootstrap
implementation details. Some of these combinations lead to bootstrap
invalidity, which we may try to resolve by another bootstrap implementation.
For example, the randomness of the bootstrap limit distribution in the
unit-root testing problem, and the associated invalidity of the standard
recursive-design bootstrap, is resolved by the use of restricted residuals
(with the unit root imposed) in the bootstrap algorithm. The challenging
part of our work is to show that the bootstrap distribution matches a
particular conditional distribution of the original statistic, in our case
conditional on the volatility process $\{\sigma _{t}\}_{t=1}^{n}$. If we
were to change the bootstrap scheme (e.g., by using an i.i.d.~or `$m$ out of 
$n$' bootstrap), then such a match would not occur, and the bootstrap would
be invalid.

Our results can be generalized in several directions. First, our
applications deal with univariate time series models and it is naturally of
interest to apply our results to multivariate (time series)\ models, where
volatilities and correlations are time-varying, stochastic and
non-stationary. In particular, in \cite{BCRT16} the bootstrap was considered
for multivariate cointegrated vector autoregressions in the presence of
stationary volatility, in combination with possible deterministic changes in
the volatility; we conjecture that our results obtained here also apply to
the case of non-stationary multivariate stochastic volatility. Second, it
would be important to understand how to bootstrap conditional mean time
series models in the presence of leverage. Although, as we have shown, the
wild bootstrap is not valid in this context, our theory may be useful for
assessing validity of other bootstrap methods when the volatility displays
leverage effects.

\section*{Acknowledgements}

We thank the guest Co-Editor and two anonymous referees for important suggestions on an earlier draft of the paper. 
We also thank participants of the conference in honour of Pierre Perron (Boston University, 14--15 March 2019),
of the 12th World Congress of the Econometric Society (Bocconi University, 17--21 August, 2020),
and of workshops and seminars in Amsterdam, Canterbury, \'{I}lhavo, Nottingham, Oxford, Rotterdam and Vienna
for useful feedbacks on earlier versions of the paper.
This research was supported by the Danish Council for Independent Research
(DSF Grant 015-00028B), by the University of Bologna (ALMA IDEA 2017 Grant),
and by the Italian Ministry of University and Research (PRIN 2017 Grant 2017TA7TYC).


\begin{thebibliography}{Cavaliere \emph{et al.}(2010a)}
\bibitem[Andrews(1997)]{A97} Andrews, D.\ W.\ K.\ (1997), \textquotedblleft
A Conditional Kolmogorov Test\textquotedblright ,\ \emph{Econometrica} 65,
1097--1128.

\bibitem[Andrews(2000)]{A00} Andrews, D.\ W.\ K.\ (2000), \textquotedblleft
Inconsistency of the Bootstrap When a Parameter Is on the Boundary of the
Parameter Space\textquotedblright ,\ \emph{Econometrica} 68, 399--405.

\bibitem[Basawa et al.(1991)]{B91} Basawa, I., A.\ Mallik, W.\ McCormick,
J.\ Reeves and R.\ Taylor (1991), \textquotedblleft Bootstrapping Unstable
First-Order Autoregressive Processes\textquotedblright ,\ \emph{The Annals
of Statistics} 19, 1098--1101.

\bibitem[Boswijk \emph{et al.}(2016)]{BCRT16} Boswijk, H.\ P., G.\
Cavaliere, A.\ Rahbek and A.\ M.\ R.\ Taylor (2016), \textquotedblleft
Inference on Co-Integration Parameters in Heteroskedastic Vector
Autoregressions\textquotedblright ,\ \emph{Journal of Econometrics} 192,
64-85.

\bibitem[Casini and Perron(2019)]{CP19} Casini, A.\ and P.\ Perron (2019),
\textquotedblleft Structural Breaks in Time Series\textquotedblright ,
in \emph{Oxford Research Encyclopedia of Economics and Finance}, Oxford:
Oxford University Press.

\bibitem[Cavaliere and Georgiev(2020)]{CG20} Cavaliere, G.\ and I.\ Georgiev
(2020), \textquotedblleft Inference Under Random Limit Bootstrap
Measures\textquotedblright , \emph{Econometrica} 88, 2547--2574.

\bibitem[Cavaliere \emph{et al.}(2015)]{CNR15} Cavaliere, G., H.\ B.\
Nielsen and A.\ Rahbek (2015), \textquotedblleft Bootstrap Testing of
Hypotheses on Cointegration Relations in VAR Models\textquotedblright ,\ 
\emph{Econometrica} 83, 813--831.

\bibitem[Cavaliere \emph{et al.}(2017)]{CNR17} Cavaliere, G., H.\ B.\
Nielsen and A.\ Rahbek (2017), \textquotedblleft On the Consistency of
Bootstrap Testing for a Parameter on the Boundary of the Parameter
Space\textquotedblright ,\ \emph{Journal of Time Series Analysis} 38,
513--534.

\bibitem[Cavaliere \emph{et al.}(2010a)]{CRT10a} Cavaliere, G., A.\ Rahbek
and A.\ M.\ R.\ Taylor (2010a), \textquotedblleft Testing for Co-Integration
in Vector Autoregressions with Non-Stationary Volatility\textquotedblright
,\ \emph{Journal of Econometrics} 158, 7--24.

\bibitem[Cavaliere \emph{et al.}(2010b)]{CRT10b} Cavaliere, G., A.\ Rahbek
and A.\ M.\ R.\ Taylor (2010b), \textquotedblleft Cointegration Rank Testing
under Conditional Heteroskedasticity\textquotedblright ,\ \emph{Econometric
Theory} 26, 1719--1760.

\bibitem[Cavaliere and Taylor(2007)]{CT07} Cavaliere, G.\ and A.\ M.\ R.\
Taylor (2007), \textquotedblleft Testing for Unit Roots in Time Series
Models with Non-Stationary Volatility\textquotedblright ,\ \emph{Journal of
Econometrics} 140, 919--947.

\bibitem[Cavaliere and Taylor(2008)]{CT08} Cavaliere, G.\ and A.\ M.\ R.\
Taylor (2008), \textquotedblleft Bootstrap Unit Root Tests for Time Series
with Nonstationary Volatility\textquotedblright ,\ \emph{Econometric Theory}
24, 43--71.

\bibitem[Cavaliere and Taylor(2009)]{CT09} Cavaliere, G.\ and A.\ M.\ R.\
Taylor (2009), \textquotedblleft Heteroskedastic Time Series with a Unit
Root\textquotedblright ,\ \emph{Econometric Theory} 25, 1228--1276.

\bibitem[Chan and Wei(1988)]{CW88} Chan, N.\ H.\ and C.\ Z.\ Wei (1988),
\textquotedblleft Limiting Distributions of Least Squares Estimates of
Unstable Autoregressive Processes,\textquotedblright\ \emph{Annals of
Statistics} 16, 367--401.

\bibitem[Crimaldi and Pratelli(2005)]{CP05} Crimaldi, I.\ and L.\ Pratelli
(2005), \textquotedblleft Convergence Results for Conditional
Expectations\textquotedblright ,\ \emph{Bernoulli} 11, 737--745.

\bibitem[Deng and Perron(2008)]{DP08} Deng, A.\ and P.\ Perron (2008),
\textquotedblleft A Non-Local Perspective on the Power Properties of the
CUSUM and CUSUM of Squares Tests for Structural Change\textquotedblright ,\ 
\emph{Journal of Econometrics} 142, 212--240.

\bibitem[Dovonon \emph{et al.}(2013)]{DGM13} Dovonon, P., S.~Gon\c{c}alves
and N.~Meddahi (2013), \textquotedblleft Boostrapping Realized Multivariate
Volatility Measures,\textquotedblright\ \emph{Journal of Econometrics} 172,
49--65.

\bibitem[Engle and Rangel(2008)]{ER08} Engle, R.\ F.\ and J.\ G.\ Rangel
(2008), \textquotedblleft The Spline-GARCH Model for Low-Frequency
Volatility and Its Global Macroeconomic Causes\textquotedblright ,\ \emph{%
The Review of Financial Studies} 21, 1187--1222.

\bibitem[Georgiev(2008)]{G08} Georgiev, I.\ (2008), \textquotedblleft
Asymptotics for Cointegrated Processes with Infrequent Stochastic Level
Shifts and Outliers\textquotedblright ,\ \emph{Econometric Theory} 24,
587--615.

\bibitem[Goggin(1994)]{G94} Goggin, E.\ M.\ (1994), \textquotedblleft
Convergence in Distribution of Conditional Expectations\textquotedblright ,\ 
\emph{Annals of Probability} 22, 1079--1114.

\bibitem[Gon\c{c}alves and Kilian(2004)]{GK04} Gon\c{c}alves, S.\ and L.\
Kilian (2004), \textquotedblleft Bootstrapping Autoregressions with
Conditional Heteroskedasticity of Unknown Form\textquotedblright ,\ \emph{%
Journal of Econometrics} 123, 89--120.

\bibitem[Gon\c{c}alves and Kilian(2007)]{GK07} Gon\c{c}alves, S.\ and L.\
Kilian (2007), \textquotedblleft Asymptotic and Bootstrap Inference for AR($%
\infty $) Processes with Conditional Heteroskedasticity\textquotedblright ,\ 
\emph{Econometric Reviews} 26, 609--641.

\bibitem[Hansen(1995)]{H95} Hansen, B.\ E.\ (1995), \textquotedblleft
Regression with Nonstationary Volatility\textquotedblright ,\ \emph{%
Econometrica} 63, 1113--1132.

\bibitem[Harvey \emph{et al.}(2016)]{HLST16} Harvey D., S.\ J.\ Leybourne,
R.\ Sollis and A.\ M.\ R.\ Taylor (2016), \textquotedblleft Tests for
Explosive Financial Bubbles in the Presence of Non-Stationary
Volatility\textquotedblright, \emph{Journal of Empirical Finance }38,
548--574.

\bibitem[H\"{a}usler and Luschgy(2015)]{HL15} H\"{a}usler, E.\ and H.\
Luschgy (2015), \emph{Stable Convergence and Stable Limit Theorems}. Cham:
Springer.

\bibitem[Jacod and Shiryaev(2003)]{JS03} Jacod, J.\ and A.\ N.\ Shiryaev
(2003), \emph{Limit Theorems for Stochastic Processes} (2nd ed.). Berlin:
Springer.

\bibitem[Kallenberg(1997)]{K97} Kallenberg, O.\ (1997), \emph{Foundations of
Modern Probability}. New York: Springer.

\bibitem[Kim and Nelson(1999)]{KN99} Kim, C.\ and C.\ Nelson (1999),
\textquotedblleft Has The U.S. Economy Become More Stable? A Bayesian
Approach Based On A Markov-Switching Model Of The Business
Cycle\textquotedblright ,\ \emph{The Review of Economics and Statistics }81,
608--616.

\bibitem[Knight(1989)]{K89} Knight, K.\ (1989) \textquotedblleft On the
Bootstrap of the Sample Mean in the Infinite Variance Case\textquotedblright
,\ \emph{The Annals of Statistics} 17, 1168--1175.

\bibitem[Lifshits(1982)]{L82} Lifshits, M.\ A.\ (1982) \textquotedblleft On
the Absolute Continuity of Distributions of Functionals of Random
Processes\textquotedblright ,\ \emph{Theory of Probability and Its
Applications} 27, 600--607.

\bibitem[Loretan and Phillips(1994)]{LP94} Loretan, M.\ and P.\ C.\ B.\
Phillips (1994),\textquotedblleft Testing the Covariance Stationarity of
Heavy-Tailed Time Series: An Overview of the Theory with Applications to
Several Financial Datasets\textquotedblright ,\ \emph{Journal of Empirical
Finance} 1, 211--248.

\bibitem[McConnell and Perez-Quiroz(2000)]{MP00} McConnell, M.\ M.\ , and
G.\ Perez-Quiros (2000), \textquotedblleft Output Fluctuations in the United
States: What Has Changed since the Early 1980's?\textquotedblright , \emph{%
American Economic Review} 90, 1464--1476.

\bibitem[Meijer(2000)]{M00} Meijer, E.\ (2000), \textquotedblleft An
Asymmetric Distribution with Zero Skewness,\textquotedblright\ Working
paper, University of Groningen, \texttt{%
http://dx.doi.org/10.2139/ssrn.2531847}.

\bibitem[Nelson(1990)]{N90} Nelson, D.\ B.\ (1990), \textquotedblleft ARCH
Models as Diffusion Approximations\textquotedblright ,\ \emph{Journal of
Econometrics} 45, 7--38.

\bibitem[Perron(2006)]{P06} Perron, P.\ (2006) \textquotedblleft Dealing
with Structural Breaks\textquotedblright , in Palgrave Handbook of
Econometrics, Vol. 1: Econometric Theory, K.\ Patterson and T.\ C.\ Mills
(eds.), Palgrave Macmillan, 2006, 278--352.

\bibitem[Ploberger and Kr\"{a}mer(1992)]{PK92} Ploberger, W.\ and W.\ Kr\"{a}%
mer (1992), \textquotedblleft The Cusum Test with OLS
Residuals\textquotedblright , \emph{Econometrica }60, 271--285.

\bibitem[Sen \emph{et al.}(2010)]{SBW10} Sen, B., M.\ Banerjee and M.\
Woodroofe (2010), \textquotedblleft Inconsistency of Bootstrap: The
Grenander Estimator\textquotedblright ,\ \emph{The Annals of Statistics }38,
1953--1977.

\bibitem[Sensier and van Dijk(2004)]{SvD04} Sensier, M.\ and D.\ van Dijk
(2004), \textquotedblleft Testing for Volatility Changes in U.S.
Macroeconomic Time Series\textquotedblright ,\ \emph{The Review of Economics
and Statistics }86, 833--839.

\bibitem[Seo(1999)]{S99} Seo, B.\ (1999), \textquotedblleft Distribution
Theory for Unit Root Tests with Conditional
Heteroskedasticity\textquotedblright ,\ \emph{Journal of Econometrics }91,
113--144.

\bibitem[Xu and Phillips(2008)]{XP08} Xu, K.-L.\ and P.\ C.\ B.\ Phillips
(2008), \textquotedblleft Adaptive Estimation of Autoregressive Models with
Time-Varying Variances\textquotedblright ,\ \emph{Journal of Econometrics}
142, 265--280.
\end{thebibliography}
\end{document}